\documentclass[prb,twocolumn,superscriptaddress]{revtex4}

\usepackage{amsmath}
\usepackage{graphicx}
\usepackage{amssymb}
\usepackage{amsthm,amscd}
\usepackage{amsfonts}
\usepackage{times}
\usepackage{pstricks}
\usepackage{wasysym}

\def\s{\sigma}

\def\tn{\textnormal}

\begin{document}

\title{Towards Universal Topological Quantum Computation in the $\nu=5/2$ Fractional Quantum Hall State}
\author{Michael Freedman}
\affiliation{Microsoft Research, Project Q, Kohn Hall, University of California, Santa Barbara, CA 93108}
\author{Chetan Nayak}
\affiliation{Microsoft Research, Project Q, Kohn Hall, University of California, Santa Barbara, CA 93108}
\affiliation{Department of Physics and Astronomy, University of California, Los Angeles, CA 90095-1547}
\author{Kevin Walker}
\affiliation{Microsoft Research, Project Q, Kohn Hall, University of California,
Santa Barbara, CA 93108}

\begin{abstract}
The Pfaffian state, which may describe the quantized Hall
plateau observed at Landau level filling fraction $\nu~=~5/2$,
can support topologically-protected qubits with extremely low
error rates. Braiding operations also allow perfect implementation
of certain unitary transformations of these qubits. However, in the case of
the Pfaffian state, this set of unitary operations is not quite sufficient
for universal quantum computation (i.e. is not dense in the
unitary group). If some topologically unprotected operations
are also used, then the Pfaffian state supports universal quantum computation,
albeit with some operations which require error correction.
On the other hand, if certain topology-changing operations can
be implemented, then fully topologically-protected universal
quantum computation is possible. In order to accomplish this, it is necessary
to measure the interference between quasiparticle trajectories
which encircle other moving trajectories in a time-dependent
Hall droplet geometry.\cite{foot}
\end{abstract}

\maketitle


\section{Introduction}

The fractional quantum Hall regime \cite{DasSarma97} contains a cornucopia
of Abelian fractional quantum Hall states, i.e. states whose quasiparticle
excitations have Abelian braiding statistics \cite{stat,Abelian-expt}.
It is possible that an even more wonderful phenomenon may
occur there: non-Abelian quantum Hall states. The strongest candidate is the
$\nu=5/2$ quantum Hall state. This state is quite
robust in the highest-mobility samples \cite{Xia04}, and
numerical studies indicate that the Pfaffian state \cite{Moore91,Greiter92},
which has excitations exhibiting non-Abelian braiding statistics
\cite{Nayak96c,Gurarie97,Tserkovnyak03,Read96,Read00,Ivanov01,Stern04,Fradkin98,Fradkin99},
has large overlap with the exact ground state for small numbers of electrons at this filling fraction
\cite{Morf98,Rezayi00}.
An experiment has been proposed \cite{DasSarma05} which would determine if the $\nu=5/2$
state is, indeed, in the universality class of the Pfaffian state by observing
the signature of non-Abelian statistics: a degenerate set of multi-quasiparticle
states which cannot be distinguished locally but can be distinguished by
a non-Abelian analogue of the Aharonov-Bohm interference measurement.
In the two quasiparticle case, this is simply the observation of a
topologically-protected qubit \cite{Kitaev97,Freedman01a}.
In the presence of many pairs of quasiparticles,
all kept far apart, the topologically-degenerate ground states form many qubits.
These states are locally indistinguishable. If the environment interacts only
locally with the system, it cannot act on these qubits.
Braiding the quasiparticles around each other, an intrinsically non-local
operation, transforms the qubits.
These gates are exact because small deformations of the
qusiparticle trajectories do not affect their braiding topology.
However, there is a sense in which the
Pfaffian state is not quite non-Abelian enough: the set of all
possible braiding operations only gives a finite set of unitary transformations
on the qubits. Thus, with these operations,
it is not possible to perform any desired unitary transformation,
which would be necessary for a universal quantum
computer.

In this paper, we suggest ways in which this apparent shortcoming
of the Pfaffian state (and, by implication, the $\nu=5/2$ quantum Hall state)
can be circumvented. The first, more pedestrian, approach is to use some non-topological
operations. Consider, for instance, what happens when two
quasiparticles are brought close together. The degeneracy between
the two states of their qubit is broken. Since unitary evolution in time will
now cause a phase difference to develop between the two states of
the qubit, we thereby implement a phase gate.
We explain how a universal quantum computer can be constructed using these ideas.
The second, more interesting, approach relies on (1) the construction by
Bravyi and Kitaev \cite{BK} of a universal set of gates for the Pfaffian state which
exploits topology change in an abstract context in which there are no restrictions
on the global topology of spacetime (using, for instance, overcrossings
and undercrossings, which seem unlikely to be realized in a system
of electrons confined to a plane ); and
(2) the observation that their operations actually {\it can} be implemented
in a way that remains entirely in the plane so long as one is able to measure the interference between
trajectories encircling quasiparticles which are moving, merging, and splitting --
i.e. interference in a time-dependent background Hall fluid.
We note, in passing, that $\nu=12/5$ may also be a non-Abelian state,
specifically one of the states proposed by Read and Rezayi \cite{Read99}.
It may be particularly interesting -- even though it is seen more weakly --
because, if it is indeed a Read-Rezayi state \cite{Read99}, braiding operations alone
are sufficient to implement any unitary transformation within
desired accuracy -- i.e. it supports topologically-protected
universal quantum computation \cite{Freedman02}. In the event
that the $\nu=12/5$ state proves to be simply an Abelian state
or to have too small an energy gap to permit manipulation, the
protocols described in this paper, if they can be experimentally realized,
would save the day by boosting the computational power of the $\nu=5/2$ state
so that it, too, can be regarded as a universal quantum computer.
Furthermore, the basic architecture which we describe in sections \ref{sec:qubit}
and \ref{sec:braiding-interfero} is relevant to the
$\nu=12/5$ state as well.

\section{Qubits in the Pfaffian State}
\label{sec:qubit}

In this paper, we will assume that the $\nu=5/2$ plateau is
in the universality class of the Pfaffian quantum Hall state.
In this section, we list some basic facts about the
Pfaffian quantum Hall state and introduce some notation
which will be useful in the following sections. The goal is to
describe the qubits which arise when many quasiparticles
\footnote{It is easier to write down explicit wavefunctions for quasiholes
than for qusiparticles, so we restrict attention in this section to quasiholes.
For the particular device configurations which we discuss in this
paper, it is also easier to work with quasiholes. However, the underlying
topological properties are the same, so we will use the terms `quasiparticle'
and `quasihole' almost interchangeably.} are present.

The Pfaffian wavefunction \cite{Moore91,Greiter92} takes the form:
\begin{equation}
\Psi_{\rm g.s.} (z_j) ~=~ \prod_{j<k} (z_j - z_k)^2 \prod_j e^{- |z_j|^2/4 }
  \cdot {\rm Pf}\!\left( \frac{1}{z_j - z_k }\right)~.
\label{grdstate}
\end{equation}
where the Pfaffian is the square root of the determinant of
an antisymmetric matrix. It has Landau level filling factor $1/2$.
(There is an obvious generalization to other even filling
factors and also to odd filling factors of bosonic particles.)
it does not appear to be a good description of electrons at
filling fraction $1/2$, which are in a metallic state down to
the lowest observable temperatures (see ref. \onlinecite{DasSarma97}
and references therein).
However, it is a candidate for the half-filled first excited Landau
level of the observed $\nu=\frac{5}{2}=2+\frac{1}{2}$ quantum
hall plateau \cite{Xia04}. If we assume that the filled lowest Landau
level of both spins is inert and translate the
Pfaffian wavefunction to the first excited Landau
level, it has high overlap with the exact ground state wavefunction
of small systems of electrons interacting through Coulomb interactions
in a half-filled first excited Landau level \cite{Morf98,Rezayi00}.
The Pfaffian state is also the exact ground state of a certain
three-body Hamiltonian \cite{Greiter92}. While this three-body Hamiltonian
is unrealistic, it has the advantage that we can also write down exact
multi-quasihole wavefunctions. Since it appears from numerical studies of
small systems that this three-body Hamiltonian is in the same universality
class as the actual Hamiltonian of the real system, we will assume that these
multi-quasihole states capture the essential topological features of
the excitations of the $\nu=5/2$ quantum Hall state.

The form of the Pfaffian factor in this wavefunction
\begin{equation}
{\rm Pf}\!\left( \frac{1}{z_j - z_k }\right) =
{\cal A}\left(\frac{1}{{z_1}-{z_2}}\frac{1}{{z_3}-{z_4}}\ldots\right)
\end{equation}
is strongly reminiscent of the real-space form of the BCS wavefunction.
Indeed, the Pfaffian state may be viewed as a quantum Hall state
of $p$-wave paired fermions \cite{Read00,Ivanov01,Stern04}.

The fundamental quasiparticles in this state carry half of
a flux quantum and, therefore, charge $e/4$. A wavefunction
for a two-quasihole state may be written as follows:
\begin{multline}
\Psi_{\rm g.s.} (z_j) ~=~ \prod_{j<k} (z_j - z_k)^2 \prod_j e^{- |z_j|^2/4 }\times\\
 {\rm Pf}\!\left(
  \frac{\left({z_j}-{\eta_1}\right)\left({z_k}-{\eta_2}\right)+{z_j}\leftrightarrow{z_k}}{z_j - z_k}\right)~.
\end{multline}
When the two quasiholes at $\eta_1$ and $\eta_2$ are brought together at
the point $\eta$, a single flux quantum quasiparticle results:
\begin{multline}
\Psi_{\rm g.s.} (z_j) ~=~ \prod_{j<k} (z_j - z_k)^2 \prod_j e^{- |z_j|^2/4 }\times\\
{\prod_i} \left({z_i}-\eta\right)\:
 {\rm Pf}\!\left(\frac{1}{z_j - z_k}\right)~.
\end{multline}

The situation becomes more interesting when we consider states
with 4 quasiholes. A wavefunction with four quasiholes at
$\eta_1$, $\eta_2$, $\eta_3$, $\eta_4$ takes the form
\begin{multline}
{\Psi_{(13)(24)}}(z_j)= \prod_{j<k} (z_j - z_k)^2 \prod_j e^{- |z_j|^2/4 }
\,\times\\
{\rm  Pf}\!\left( \frac{(z_j - {\eta_1}) (z_j - {\eta_3} )(z_k
-{\eta_2} ) (z_k - {\eta_4} ) + (j \leftrightarrow k )}
{z_j - z_k}\right)
\label{qhwf}
\end{multline}
However, there is another wavefunction with four quasiholes at
$\eta_1$, $\eta_2$, $\eta_3$, $\eta_4$:
\begin{multline}
{\Psi_{(14)(23)}}(z_j)= \prod_{j<k} (z_j - z_k)^2 \prod_j e^{- |z_j|^2/4 }
\,\times\\
{\rm  Pf}\!\left( \frac{(z_j - {\eta_1}) (z_j - {\eta_4} )(z_k
-{\eta_2} ) (z_k - {\eta_3} ) + (j \leftrightarrow k )}{z_j - z_k}\right)
\label{qhwf-2}
\end{multline}
These two wavefunctions are linearly independent, but they have the same charge
density profiles so long as $\eta_1$, $\eta_2$, $\eta_3$, $\eta_4$
are far apart. In fact, they are indistinguishable by any local measurement.
One might even think that there is a third four-quasihole state ${\Psi_{(12)(34)}}$,
but this is not independent of the other two\cite{Nayak96c},
\begin{equation}
{\Psi_{(14)(23)}}-{\Psi_{(12)(34)}}=x\left({\Psi_{(13)(24)}}-{\Psi_{(12)(34)}}\right)
\end{equation}
where $x=\left({\eta_1}-{\eta_3}\right)\left({\eta_2}-{\eta_4}\right)/
\left({\eta_1}-{\eta_4}\right)\left({\eta_2}-{\eta_3}\right)$.

It is enlightening to pick as the basis of the two dimensional space
of four-quasihole wavefunctions ${\Psi_{(13)(24)}}$ and the following state
\cite{Read96} (${\cal N}$ is a normalization factor):
\begin{multline}
{\Psi_{(13)(24)}}-{\Psi_{(13)(24)}}={\cal N} \prod_{j<k} (z_j - z_k)^2 \prod_j e^{- |z_j|^2/4 }
\,\times\\
{\cal A }\!\Biggl( {z_1^0} {z_2^1}\,\frac{(z_3 - {\eta_1}) (z_3 - {\eta_3} )(z_4
-{\eta_2} ) (z_4 - {\eta_4} )}
{z_3 - z_4} \\
\frac{ (z_5 - {\eta_1}) (z_5 - {\eta_3} )(z_4
-{\eta_2} ) (z_6 - {\eta_4} )}{z_5 - z_6}
\ldots\Biggr)
\label{qhwf-nf}
\end{multline}
The interpretation is that a Cooper pair can be broken and the resulting
neutral fermions put into zero modes\cite{Read96}. In this
case, the zero modes have wavefunctions $z^0$ and $z^1$; in
the $n$-quasihole case, they have the form $z^k$ with $0\leq k \leq n-1$.
In fact, a Cooper pair can even be broken when there are only two quasiparticles,
but only one of the neutral fermions can go into a bulk zero mode;
the other one must be at the edge. (Similarly, in the four-quasihole
case, there are two additional states in which a neutral fermion
is in one of the two zero modes while the other one is at the edge.)

In this way, it can be seen that there are $2^{n}$
$2n$-quasihole states \cite{Nayak96c,Read96} (half
of them have a neutral fermion at the edge and the other half
don't). It has been shown \cite{Read00}
that precisely the same degeneracy is obtained in a $p~+~ip$
superconductor when there are $2n$ flux $hc/2e$ vortices
present: there is one zero mode solution of
the Bogoliubov-De Gennes equations per vortex. These solutions
are Majorana modes. Grouping them into pairs, we have $n$
fermionic levels, each of which can be occupied or unoccupied.
By breaking Cooper pairs, we can change their occupancies.
{\it We interpret this degeneracy as $n$ qubits, one qubit
for each pair of quasiholes}.
(Of course, the grouping of quasiholes into pairs
is arbitrary and any two pairings are related by the a change of basis.)

\begin{figure}[tbh]
\includegraphics[width=3.5in]{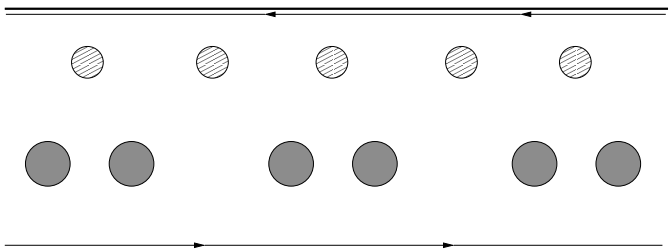}
\caption{A system with $n$ quasihole pairs (held at pairs of anti-dots,
depicted as shaded circles) supports $n$ qubits. Additional antidots
(hatched) can be used to move the quasiparticles, as described in
section \ref{sec:braiding-interfero}.}
\label{fig:multi-qhs}
\end{figure}

Hence, we envision platform for quantum computation depicted in
Figure \ref{fig:multi-qhs}.
An $n$-qubit system can be created by endowing a Hall bar with
$2n$ antidots at which quasiholes are pinned. Each pair of quasiholes
has a two-dimensional Hilbert space spanned by
$|0\rangle$ and $|1\rangle$, which correspond to the absence or presence of
a neutral fermion. In the following sections, we will
discuss how these qubits can be manipulated and measured.

These qubits will be manipulated by braiding quasiparticles,
which causes states in this $2^n$-dimensional Hilbert
space transform into each other. To discuss these
transformations, a different basis than (\ref{qhwf}) is useful.
The effect of braiding quasiparticles is a combination
of the explicit monodromy of the wavefunction and the
Berry matrices obtained from adiabatic transport of the $\eta_i$s.
The phase factors in (\ref{eqn:fourqh}) below have been chosen
so that the latter are trivial and the former completely
encapsulate quasiparticle braiding properties \cite{Nayak96c}.
(We could have worked with the basis ${\Psi_{(13)(24)}}$, ${\Psi_{(14)(23)}}$,
in which there is no explicit monodromy, but then we would have to
evaluate Berry matrix integrals \cite{Gurarie97,Tserkovnyak03}.)
\begin{multline}
\label{eqn:fourqh}
{\Psi^{(0,1)}}(z_j) =  \frac{\left({\eta_{13}}{\eta_{24}}\right)^{\frac{1}{4}}}
{(1 \pm \sqrt{x})^{1/2}}\,
\left( {\Psi_{(13)(24)}} \,\,\pm\,\,\sqrt{x}\,\,
{\Psi_{(14)(23)}}\right)
\end{multline}
where $\eta_{13}={\eta_1}-{\eta_3}$, etc. and
$x~=~\eta_{14}\eta_{23}/\eta_{13}\eta_{24}$.
(Note that we have taken a slightly different
anharmonic ratio $x$ than in Ref. \onlinecite{Nayak96c}
in order to make (\ref{eqn:fourqh}) more compact than
Eqs. (7.17), (7.18) of Ref. \onlinecite{Nayak96c}.)
From this expression, we see, for instance,
that taking $\eta_3$ around $\eta_1$
transforms ${\Psi^{(0)}}(z_j)$ into ${\Psi^{(1)}}(z_j)$.

In the $2n$ quasihole case, the result can be stated as follows\cite{Nayak96c,Ivanov01}.
The $2^n$ states of the system can be grouped into
a representation of the Clifford algebra
\begin{equation}
\left\{ {\gamma_i},{\gamma_j}\right\} = 2\delta_{ij}
\end{equation}
with $i,j=1,2,\ldots,2n$. We could, for instance, organize the
states according to their eigenvalues ${\psi_i^\dagger}\psi^{}_i=\pm 1$,
where $\psi^{}_j=\gamma_{2j-1}+i\gamma_{2j}$, $j=1,2,\ldots,n$.
When quasiparticles $i$ and $j$ are exchanged,
the states transform according to\cite{Nayak96c,Ivanov01}:
\begin{equation}
|\Psi\rangle \rightarrow e^{-\frac{\pi}{4}{\gamma_i}{\gamma_j}} |\Psi\rangle
\label{eqn:braiding-rules}
\end{equation}
These braiding matrices, $\tau_{ij}=\exp\left(-{\frac{\pi}{4}{\gamma_i}{\gamma_j}}\right)$
will be a set of topologically-protected unitary transformations which
we can use to manipulate our qubits.

Several important calculational and heuristic tools follow from
field theories for the Pfaffian state. While they illuminate this
section and section \ref{sec:braiding-interfero}, they are somewhat
technical and take us away from the main line of our exposition, so we have deferred
a discussion of these field theories to appendix \ref{sec:EFT}.
For reasons which are discussed there, it is convenient to
call excitations which have the same braiding properties (up
to Abelian phase factors)
as, respectively, (a) the vacuum, (b) charge-$e/4$ quasiparticles, and (c)
neutral fermions either $1,\sigma,\psi$ or, equivalently, isospin $0,\frac{1}{2},1$.

Thus far, we have assumed that the only quasiparticles in our system
are the quasiparticles which we have induced on our anti-dots. There could
also be thermally-excited quasiparticles. They are the main source of
error and their density was estimated in ref. \onlinecite{DasSarma05} to be exponentially
small at low temperatures. However, even at zero temperature,
there will always be some quasiparticles which are trapped by local
variations in the potential, such as those cause by impurities.
Assuming that they cannot move, the effect of these quasiparticles can always be accounted
for with `software', i.e. quantum computations
must be done with some more complicated algorithms which compensate
for the presence of these stray quasiparticles. As a practical matter, however,
we would like to make them as benign as possible. To the extent that
we can tune the magnetic field to the center of the plateau and use gates
to move the edge of the system to avoid impurities (as in ref. \onlinecite{Stern05}),
we should do so. If we can remove these localized quasiparticles
with gate or a scanning probe microscope tip (such as an atomic-force microscope (AFM)
tip) we should also attempt this. Finally, there is
one simplifying feature of the Pfaffian state in particular,
noted in refs. \onlinecite{Stern05,Bonderson05} is
that there is an even-odd effect with quasiparticles. An even number of
quasiparticles fuse to form a quasiparticle with Abelian statistics,
while an odd number of quasiparticles fuse to form a quasiparticle
with non-Abelian statistics. Hence, we handle stray localized quasiparticles
in the following way. We should associate each stray quasiparticle with
one of the anti-dots (most naturally the anti-dot to which it is closest).
Then, we want the number of stray quasiparticles
associated with each anti-dot to be even. In this way, the degrees of freedom
of the anti-dot are not modified by its associates. Finally, we need to
ensure that the quasiparticle braiding trajectories always encircle
an even number of stray quasiparticles. Then, as may be seem from
repeated application of (\ref{eqn:braiding-rules}),
the braiding matrices acting on the qubit Hilbert space are unaffected by
the presence of the stray quasiparticles and the Hilbert
space of the stray quasiparticles will not become entangled with it.

\section{Braiding and Interferometry}

\label{sec:braiding-interfero}

\subsection{Braiding}

\begin{figure}[tbh]
\includegraphics[width=3.5in]{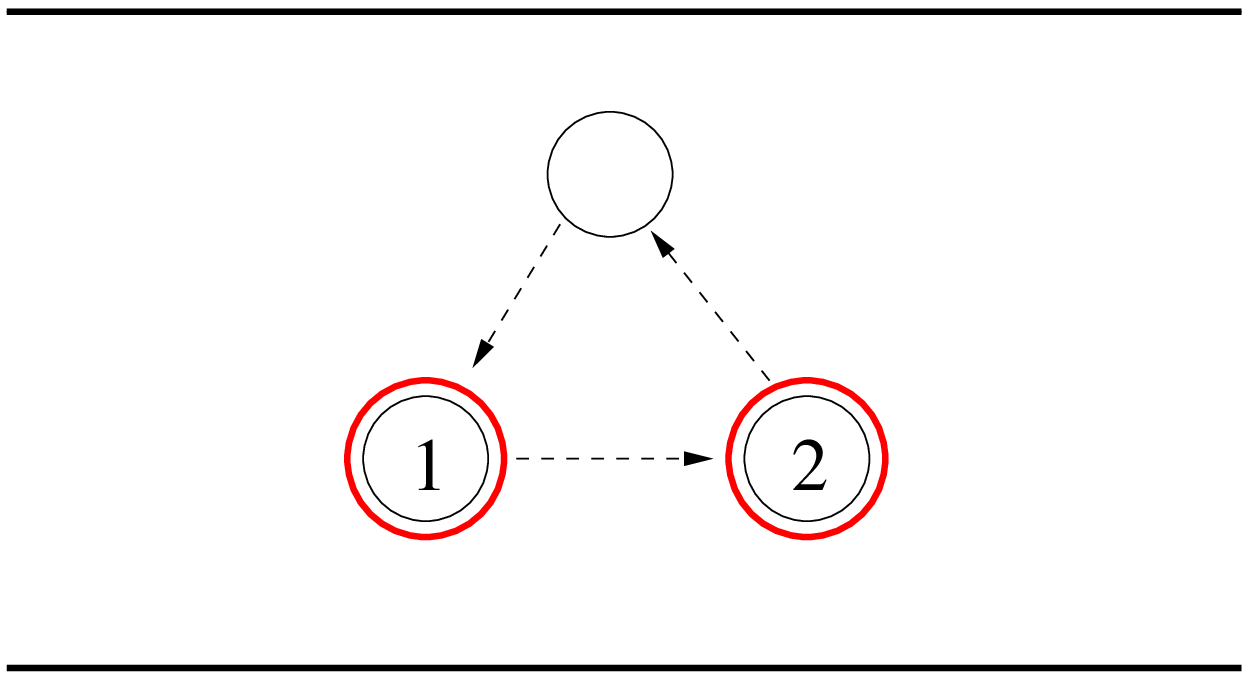}
\caption{The exchange of two qubits through a third anti-dot.}
\label{fig:exchange-pump}
\end{figure}

The basic process by which we will manipulate
many-quasiparticles states is the counterclockwise exchange
depicted in Figure \ref{fig:exchange-pump}. We suppose that the quasiparticles
are localized at anti-dots and that they are transferred from one anti-dot
to another by varying the voltages on the anti-dots. With three anti-dots,
an exchange can be performed. Two successive exchanges results in
a full braid of one quasiparticle around the other.
We may need to move one quasiparticle greater distances --
for instance, to take it around several others --
in which case we could use an array of anti-dots as
a `bucket brigade' (as in CCD devices such as digital cameras).

The process depicted in Figure \ref{fig:exchange-pump}
can be used, for instance, to
exchange a quasiparticle from
one qubit with a quasiparticle from a different qubit. Such a process,
which applies the gate $g_3$ ($g_1$ and $g_2$ will be introduced
later) has a spacetime diagram which is depicted in Figure \ref{fig:g_3}:
\begin{eqnarray}
{g_3} = \frac{1}{\sqrt{2}}\left(
 \begin{array}{cccc}
  1 & 0 & 0 &  -i \\
  0 & 1 &  -i & 0 \\
  0 & -i & 1 &  0   \\
   -i & 0 & 0 & 1
 \end{array}
\right). 
\end{eqnarray}
This will be one of the basic gates used below.

\begin{figure}[tbh]
\includegraphics[width=3.5in]{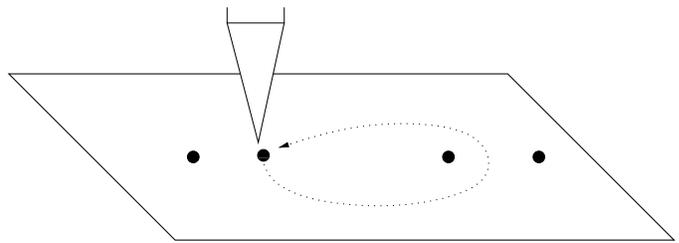}
\caption{Using an AFM tip to braid quasiparticles.}
\label{fig:AFM}
\end{figure}

One can imagine another possibility, which might become realistic at some point in the future:
one quasiparticle could be dragged around another with a scanning probe microscope
tip, e.g. an AFM tip, 
which could couple to a quasiparticle electrostatically with the
required spatial resolution (which is presumably the magnetic length,
on the order of 100 \AA),
as depicted in Figure \ref{fig:AFM}.

\begin{figure}[tbh]
\includegraphics[width=3.25in]{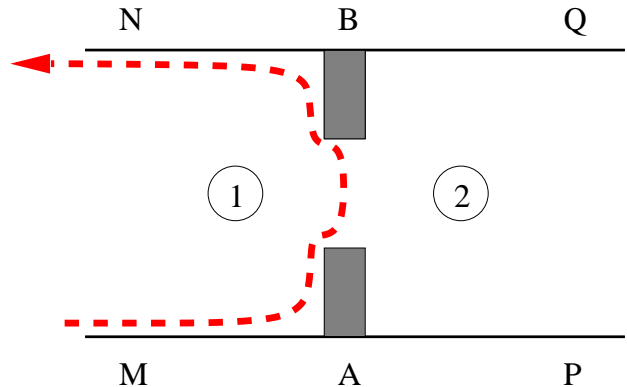}
\caption{Taking a quasiparticle from the edge around one
of the quasiparticles in a qubit implements a logical NOT
on the qubit.}
\label{fig:NOT}
\end{figure}

There is one other type of braiding process which we can do, namely
taking a quasiparticle from the edge of the system around one
of the quasiparticles in a qubit, as depicted in Figure \ref{fig:NOT}.
Such an process is a NOT gate for this qubit.

\subsection{Interferometry}

\begin{figure}[b]
\includegraphics[width=3.25in]{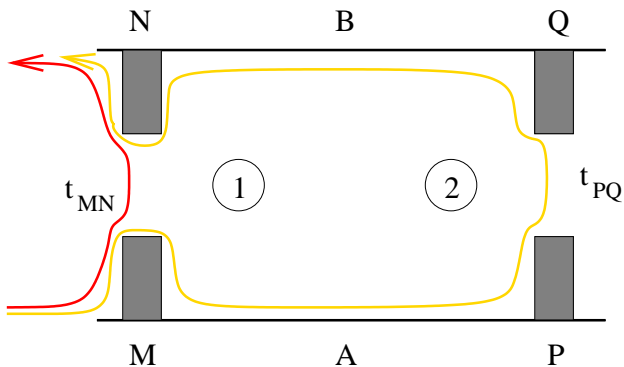}
\caption{The state of a qubit can be determined from a measurement
of the longitudinal conductance when inter-edge tunneling is allowed
at two interfering junctions.}
\label{fig:read}
\end{figure}

The basic process by which we will determine the state of our
system, i.e. read our qubits, is an interference measurement.
The two states of a qubit, which differ by the absence or presence of a
neutral fermion $\psi$, can be distinguished by taking a charge
$e/4$ quasiparticle around the pair. If the neutral fermion $\psi$ is present,
an extra $(-1)$ occurs in the amplitude. In ref. \onlinecite{DasSarma05},
it was shown how this minus sign could be detected by
measuring the longitudinal resistance, $\sigma_{xx}$.
It is determined by the probability for current entering the bottom edge from the left
in Figure \ref{fig:read} to exit along the top edge to the left.
This is given,
to lowest order in $t_{MN}$ and $t_{PQ}$, by the interference between two
processes: one in which a `test' quasiparticle tunnels from M to N;
and another in which the `test' quasiparticle instead continues along the bottom
edge to P, tunnels to Q, and then moves along the top edge to N.
(We subsume into $t_{PQ}$ the phase associated with the extra distance
travelled in the second process and the extra Aharonov-Bohm phase.)
\begin{equation}
\label{eqn:inter-cond}
\sigma_{xx} \propto |t_{MN}|^2 + |t_{PQ}|^2 +
2\text{Re}\left(t_{MN}{t_{PQ}^*} \left\langle\psi \left| B \right|\psi\right\rangle\right)
\end{equation}
The third term is the interference between the two possible
tunneling trajectories. $|\psi\rangle$ is the state of the qubit and the test
quasiparticle, and $B$ is the operator which takes the test quasiparticle
around the qubit, i.e. the braiding matrix\cite{Fradkin98}. It can be computed by
any of three equivalent ways: (1) taking
$\eta_3$ around  $\eta_1$ and $\eta_2$ in equation (\ref{eqn:fourqh}); (2)
using the expression in (\ref{eqn:braiding-rules});
or (3) by evaluating the Jones polynomial at $q=e^{\pi i/4}$ for the
link diagrams in Figure \ref{fig:interfere-link} (see appendix \ref{sec:EFT} for
more on the meaning and evaluation of these diagrams). Either of these methods
shows that $\left\langle\psi \left| B \right|\psi\right\rangle = \pm i$ for the
two states of the qubit (the factor of $i$ comes from the Abelian sector
of the theory). Hence,
\begin{equation}
\sigma_{xx}\propto |t_{MN}\pm i\,t_{PQ}|^2 .
\end{equation}
with the $+$ sign corresponding to the state $|0\rangle$ and the $-$
sign corresponding to the state $|1\rangle$.

\begin{figure}[t]
\includegraphics[width=3.25in]{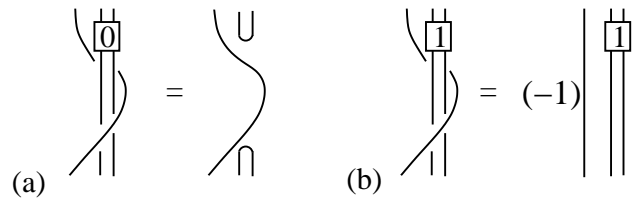}
\caption{The interference between the two trajectories in Figure \ref{fig:read}
can be obtained from the Jones polynomial (operator) evaluated on
the two diagrams in this figure. In (a) the qubit is in the state $0$,
while in (b) it is in state $1$.}
\label{fig:interfere-link}
\end{figure}

In the many-qubit device of Figure \ref{fig:multi-qhs}, we would need
tunnel junctions on either side of each qubit. By doing a sequence of
tunneling conductance measurements, we could read each qubit
in succession.
 
\begin{figure}[tbh]
\includegraphics[width=3.25in]{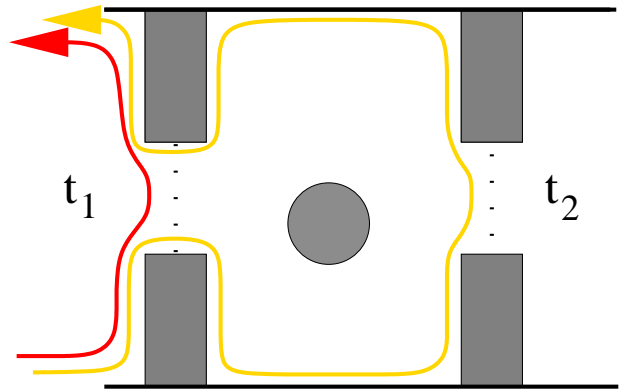}
\caption{When there is a charge $e/4$ qusiparticle at the anti-dot in the middle
of the device, then there is no interference between the two trajectories from $X$ to
$Y$ contributing to the longitudinal conductance.}
\label{fig:claudio-dev}
\end{figure}

 The presence or absence of a charge $e/4$ quasiparticle on an
 anti-dot can, of course, be detected simply by measuring the charge
 on the anti-dot. However, we will have occasion to measure the topological
 charge contained within some complicated spacetime loops, so it will
 also be useful to detect charge $e/4$ quasiparticles by interferometry.
 This can be done using the experimental setup of refs. \onlinecite{Chamon97,Fradkin98},
 as analyzed in refs. \onlinecite{Stern05,Bonderson05}. When a charge $e/4$ quasiparticle
 is present on the anti-dot in Figure \ref{fig:claudio-dev}, the authors of
 refs. \cite{Stern05,Bonderson05} showed, the two trajectories
 do not interfere at all because $\left\langle\psi \left| B \right|\psi\right\rangle =0$
 for equation (\ref{eqn:inter-cond}) applied to this
 setup. This may be seen by evaluating the Jones polynomial at
 $q=e^{\pi i/4}$ for the first link in Figure \ref{fig:Jones-qubit} (see appendix
 \ref{sec:EFT} for more details). Hence,
 $\sigma_{xx}\propto {|{t_1}|^2}+{|{t_2}|^2}$.
 Varying the phases of $t_1$ and $t_2$ will not affect the
 longitudinal conductivity, which is the signature of a $\sigma$ particle.
 
 \subsection{Tilted Interferometry}
 
\begin{figure}[bht]
\includegraphics[width=3.05in]{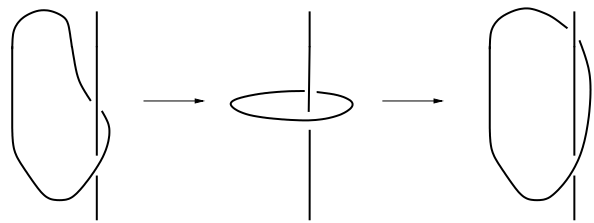}
\caption{A quasiparticle trajectory which winds around the antidot (straight line)
can be deformed into a single time slice or stretched over a long time.}
\label{fig:spacetime-slice}
\end{figure}

 We now consider a generalization of the interferometry measurements
 of the previous subsection. Consider the diagram of Figure \ref{fig:spacetime-slice},
 in which a quasiparticle-quasihole pair is created, one member of the pair
 winds around another quasiparticle fixed at an antidot, and then the pair is again annihilated.
 This picture has a special feature, namely that the quasiparticle-quasihole
 loop can be continuously deformed into a single time slice or, for that matter,
 stretched out so that it takes place over a very long time, as in the third picture in figure
 \ref{fig:spacetime-slice}. Because the
 antidot is simply sitting there passively, the evolution in the time direction 
 can be chosen to our advantage.

\begin{figure}[tbh]
\includegraphics[width=2.5in]{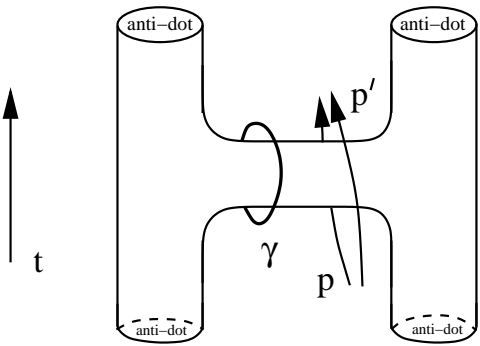}
\caption{Two antidots are merged for a short while and then separated
again. The spacetime curve $\gamma$ encircles the merger region.
The interference between the trajectories $p$ and $p'$ measure the topological
charge around this curve.}
\label{fig:spacetime-tilted}
\end{figure}

The amplitude for such a process is a measure
of the total topological charge
of the planar region bounded by this loop when
deformed into a single time slice (as in the middle picture
in Figure \ref{fig:spacetime-slice}) -- in other words, it measures the topological charge on the antidot.
However,  we are also free to consider such
processes even when they do not have an interpretation in terms of the charge
in some region of a fixed time-slice plane. This type of process can occur when the spacetime
topology is non-trivial. For instance, if the system is on a torus,
then there is a process in which a quasiparticle-quasihole pair is created, the quasiparticle
taken around the meridian of the torus until it again meets the quasihole, and
they are both annihilated. The corresponding loop does not enclose any region,
so the usual interpretation is not available. Such loops are an important part
of the Bravyi-Kitaev construction which we describe in section \ref{sec:Bravyi-Kitaev}.
 An even more exotic possibility is depicted
 in Figure \ref{fig:spacetime-tilted}. Suppose we have two antidots
 which we bring close together so that they fuse for a short
 period of time $T$ before we pull them apart again.
 The spacetime diagram for this process is depicted in figure
 \ref{fig:spacetime-tilted}. Now consider a test quasiparticle which travels between
 the two dots. We have drawn two interfering trajectories which the
 test quasiparticle can take, labeled $p$ and $p'$ in Figure \ref{fig:spacetime-tilted}.
 One of these trajectories, $p$, passes between the antidots before the merger while the
 other, $p'$, passes between the anti-dots after the merger. The curve $\gamma$
 captures the matrix element for the interference between these two trajectories.

 Ordinarily, one thinks of the amplitude of Figure \ref{fig:spacetime-tilted} as being quite
 different from the middle one in Figure \ref{fig:spacetime-slice} (for instance,
 in Yang-Mills theory, a Wilson loop in the time direction
 is a measure of the force between separated charges,
 and therefore is a probe of confinement). However, in a topological phase,
 the curve $\gamma$ in Figure \ref{fig:spacetime-tilted} is put on the same
 footing as the loop in Figure \ref{fig:spacetime-slice}
 (see appendix \ref{sec:EFT} for more on the relation between these diagrams
 and matrix elements in Chern-Simons theory).
 In a topological phase, the system does not know about any preferred
 metric (at least at long distances and low energies),
 so the time direction is just as good as a spatial direction.
 The results of such interferometry measurements correspond to
 topological charges, even though there isn't an interpretation
 as the topological charge enclosed within a spatial loop. This may be familiar
 to some readers in the context Laughlin states in the quantum Hall effect.
 At $\nu=1/3$, an interference experiment around the meridian of a torus
 can return as its answer $q=0,\frac{1}{3},\frac{2}{3}$ (modulo $1$). Of course, the
 meridian of the torus does not enclose a region, so these are not charges
 enclosed within a meridional loop. Rather, these results
 correspond to the different possible quasiparticle boundary conditions (monodromies)
 around the meridian (namely $\psi\rightarrow e^{2\pi i m/3}\,\psi$, $m=0,1,2$)
 which are the same as the phases which a quasiparticle
 would acquire in going around a region containing charges $q=0,\frac{1}{3},\frac{2}{3}$.
 Similarly, in the Pfaffian state tilted interferometry experiments will return
 as their result either $1,\sigma,\psi $, just as ordinary interferometry does.
 
 Such measurements, in which quasiparticles encircle the time-dependent
 trajectories of quasiparticles on anti-dots, will be important for the protocols proposed
 in this paper, so it is worth spending a little time determining the limitations
 on such measurements, which we will call `tilted interferometry' because
 the curve $\gamma$ cannot be deformed into a single time slice.
 (We thank Ady Stern for pointing out that tilted
 interferometry is analogous to the less well-known
 experiment proposed by Aharonov and Bohm \cite{Aharonov61},
 in which the time dependence of $A_0$ affects quasiparticle interference
 between trajectories which do not pass through regions of finite electric
 (or magnetic) field.)
 
\begin{figure}[tbh]
\includegraphics[width=3in]{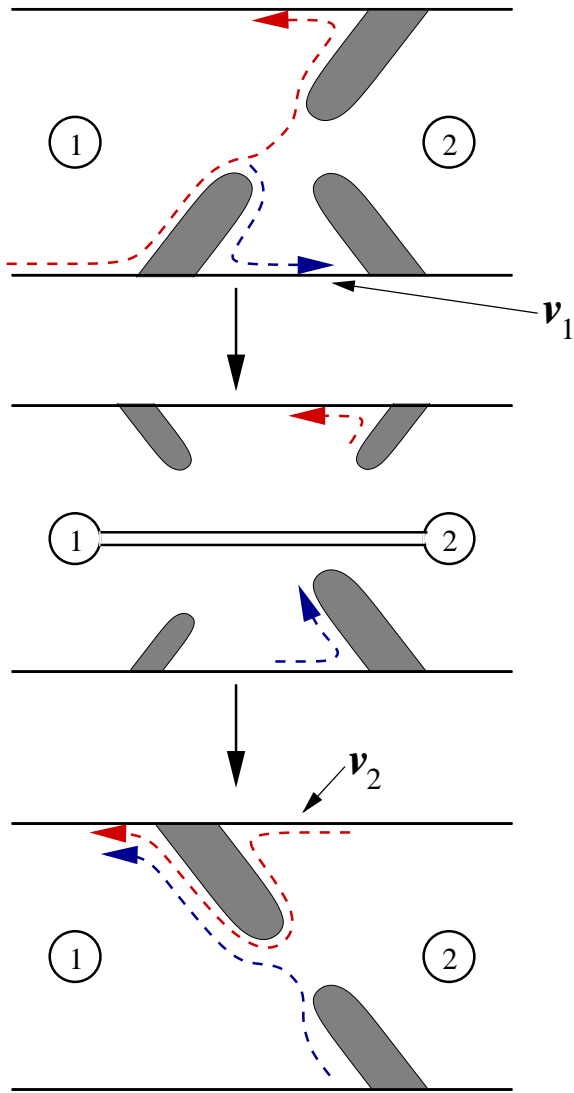}
\caption{Two antidots are merged for a short while and then separated
again. By putting delays into the possible quasiparticle trajectories (depicted in red and blue), we can enable them to interfere. In order to give the two
anti-dots enough time to merge, we might wish to make the velocities
${v_1}$, ${v_2}$ small.}
\label{fig:tilted-traj}
\end{figure}

 Unlike in the case of `ordinary' interferometry,
 a tilted measurement cannot, strictly speaking,
 be a DC measurement. In an ordinary interference experiment,
 the different interfering quasiparticle
 wavefunctions are plane-wave-like states, hence,
 even though the travel time for
 the two trajectories in Figure \ref{fig:read} will be different,
 the wavefunctions will have
 spatio-temporal overlap, so they will still be able to interfere.
 Consider, however, the trajectories in Figure \ref{fig:spacetime-tilted}.
 Let us suppose that the two anti-dots are merged from time $t_1$ until time
 $t_2$. Then the first trajectory between the anti-dots must occur before $t_1$
 while the second trajectory must occur after $t_2$. The only way that the wavefunctions for
 the two quasiparticle trajectories can have an overlap is if there is a delay built
 into the first trajectory which will allow the second one to `catch up'. This can
 be done as shown in Figure \ref{fig:tilted-traj}. We turn on and off some of the gates
 in order to direct the quasparticles along the specified trajectories.
 It will also be helpful to vary the quasiparticle velocities, as also shown in the figure.
  
 \section{Topological Protection}
 \label{sec:protection}
 
The main advantage of using a topological state as a platform
for quantum computation is that
such states have intrinsic fault-tolerance \cite{Kitaev97}.
The multi-quasihole states cannot be distinguished by local
measurements, so long as the quasiholes are kept far apart.
Hence, interactions with the environment,
which are presumably local, cannot cause transitions between different topologically-degenerate
states nor can it split them in energy.
Suppose that we have $2n$ quasiparticles in our system,
which is the device of Figure \ref{fig:multi-qhs}.
One might worry, for instance, that there could be a local voltage fluctuation
at one of the anti-dots.
This has a trivial effect, however: the voltage fluctuation changes the energy of all $2^n$ states of our Hilbert space by precisely the same amount. Thus,
it does not apply a phase gate, as could happen in a non-topological
quantum computing scheme.
The only way in which errors can occur is
if a stray quasiparticle (created by the interaction with the environment)
moves across the system and spontaneously performs a topological operation such as the braiding operation of Figure \ref{fig:NOT} or the interference measurement of Figure \ref{fig:read}.
This is closely related to the longitudinal resistance, which
is the probability of an event in which a quasiparticle travels
from one end of the system to another. If the spacing
between anti-dots is large enough that quasiparticle
transport at this scale is in the ohmic regime, then the error rate
and the longitudinal resistance will be
controlled by the same kind of processes. (We thank Leon
Balents for a discussion of this point.)
Experimentally, the longitudinal resistance is
observed to be in the thermally-activated regime.
Hence, it is limited by the density of thermally-excited quasiparticles,
which is exponentially small $\exp(-\Delta/{k_B}T)$, where $\Delta$ is
the quasiparticle energy gap and $T$ is the temperature.
\footnote{Presumably it is controlled by variable-range hopping
at still lower temperatures. In this regime, a quasiparticle which
is in one localized state at zero temperature is thermally
excited into another localized state. Through a sequence
of such hops, with an average range which varies with temperature,
it travels across the system.}
In ref. \onlinecite{DasSarma05}, the resulting error rate $\Gamma$
was estimated to be
\begin{equation}
\frac{\Gamma}{\Delta} \sim \frac{{k_B}T}{\Delta} \, e^{-\Delta/{k_B}T}
\end{equation}
In order to minimize
the error rate, we want the temperature to be as low as possible,
and we want the gap to be as large as possible, which seems to be aided
by ultra-high mobility samples.
The lowest temperature reached in the experiment of ref. \onlinecite{Xia04}
was $T=5\, {\rm mK}$, while the measured gap was
$\Delta/{k_B}~=~0.5 {\rm K}$.
This leads to an error rate less than $10^{-30}$.
 
 Of course, quasiparticles cannot be kept infinitely far apart,
 so there will be some splitting between multi-quasihole states.
 This splitting can be understood as the formation of a band of
 propagating Majorana fermion modes, which mix the localized
 states. The width of this band will be proportional to the tunneling
 matrix element between two quasiholes, which should decay
 as $w\sim e^{-R\Delta/c}$, for some constant $c$ with dimensions of velocity,
 where $R$ is the distance between quasiholes and $\Delta$ is the quasiparticle
 gap. The condition that the quasiparticles
 should be kept far apart can be translated into the statement that
 braiding operations should be done on time scales shorter
 than $1/w$. By keeping $R$ large compared to the inverse of the gap
 we can ensure that this will always be the case.
 
 When quasiparticles are brought close together, however,
 there is no longer exponential protection. Suppose, for instance,
 that we merge two antidots into one large antidot of radius $\rho$.
 The splitting between the states $|0\rangle$ and $|1\rangle$
 is now determined by processes in which a quasiparticle-quasihole
 pair is created at the edge of the antidot, they move in opposite directions
 around the antidot, and annihilate on the other side. Since there is
 no gap for the creation of quasiparticles at the edge of the antidot,
 there is no longer exponential suppression of such a process. Instead,
 it leads to a splitting $w \sim 1/\rho$. This means
 that the resulting phase error will be small if two quasiparticles are
 merged into a large antidot for a short period of time,
 even though the protection is not as good as exponential.
 However, there is still topological protection against bit flip errors
 since these would require a neutral fermion to tunnel to or from the qubit.
 
 A second aspect of topological protection is the exactness
 of braiding operations. In the case of, say, spin qubits, gates are
 necessarily noisy because they depend on our ability to
 precisely tune the duration of a $\pi$ pulse or the strength of an
 applied magnetic field, which is necessarily imperfect. When gates are
 applied by braiding quasiparticles, however, no such tuning
 is necessary. The process is discrete: we either braid two quasiparticles
 or we don't, and if we do braid them, then the corresponding
 unitary transformation occurs with the same level of exactness
 as the vanishing of the longitudinal resistivity or the quantization
 of the Hall resistivity.
 
\begin{figure}[t]
\includegraphics[width=3.25in]{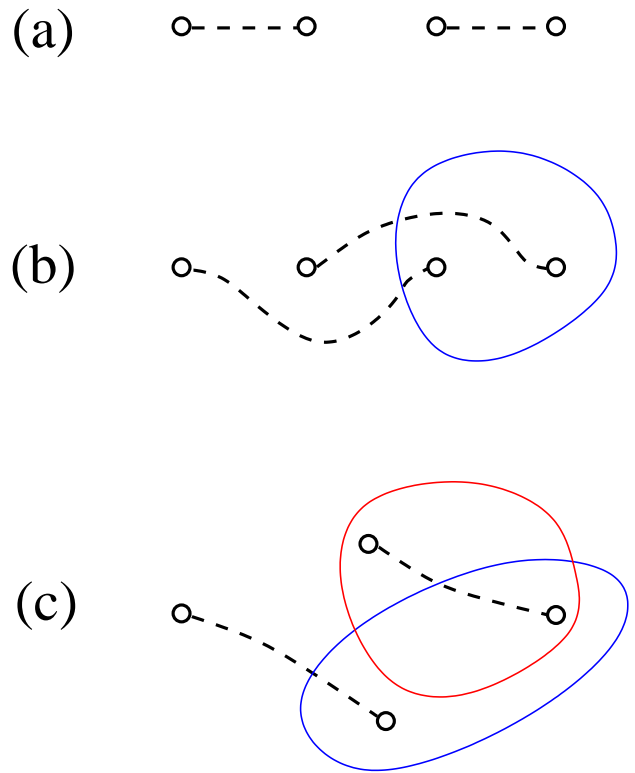}
\caption{In (a), we have two pairs of quasiparticles. The
dashed lines connect quasiparticles which fuse to
form the (topologically) trivial particle. In (b), the
second and third quasiparticles are exchanged.
A measurement of the topological charge around the
blue curve or an operation which takes another quasiparticle
around this curve will give a non-trivial result as a result of 
this exchange. The same result is clearly obtained even if the
final position of the quasiparticles is slightly different from
those shown. If the exchange is incomplete, as in
(c), then a measurement around the blue curve will
give a non-trivial result but a measurement around the red curve
will give a trivial result. So long as we are careful to measure the
system with the blue curves in (b) and (c), we will find the correct
result.}
\label{fig:partial-braid}
\end{figure}

However, one might wonder what happens if a quasiparticle only
goes 359 degrees around another. If our qubits were quasiparticle
pairs which we created out of the vacuum and then measured after
annihilating them again in pairs at the end of the computation, then it
would be clear that we would be dealing with closed braids. So long
as the topological class of the closed braid traced out by
the entire history of the system were preserved,
it would not matter whether one quasiparticle went 360 degrees
around another or only part of the way around. However,
we envision measuring our qubits through a quantum interference
measurement of the topological charge around some closed curves.
Therefore, we will consider this issue in a little more detail.
If one quasiparticle were to go 360 degrees
around another then the initial and final states of the
system lie in the same Hilbert space and the action of the braiding
operation is just its unitary representative on this Hilbert space.
However, if a particle only goes 359 degrees around another,
then the initial and final states of the system do not lie in the same
Hilbert space. Of course, the initial and final Hilbert spaces are unitarily
equivalent, but the problem is that a unitary transformation between
them could be trivial or it could undo the braiding operation.
So what happens?
The answer is that it depends on how the system
is now measured. If the state of the system is measured by interference,
then the result will depend on what path the interfering
test quasiparticle takes (incidentally, this is always true since we
can undo the effect of a braid by choosing a convoluted interfering
path). If a particle goes 359 degrees around another, then almost all
paths will give a result which is the same as if it had gone 360 degrees.
In other words, it is not necessary to have very precise
control of quasiparticle positions. Of course,
if a quasiparticle were to only go 270 degrees around another,
then we would have to exercise more care in choosing a trajectory for
a test quasiparticle. However, even as drastic a deviation as this
is not that serious a problem.
The same caution holds if, instead of measuring the
system, we wanted to act on it with yet another braiding operation.
An example of such a situation is given in Figure \ref{fig:partial-braid}.
The same logic also holds for the uncontrolled motion of a stray
quasiparticle; whether or not it has moved far enough to cause an error
will depend on how the qubit is subsequently measured.

 A potential second source of error is braiding
 operations which are performed
 too quickly. The time scale $t_{\rm op}$ over which a braiding operation is
 done must be slow compared to the gap, $t_{\rm op} \gg \Delta$.
 If this inequality is violated, a pair of quasiparticles might be created.
 These quasiparticles might then execute a non-trivial braid before
 annihilating each other (Coulomb blockade presumably prevents them
 from annihilating the quasiparticles on the anti-dots), thereby applying
 $g\cdot g'$, for some $g'$, rather than our intended gate $g$. The amplitudes
 for various $g'$ depend on the ability of quasiparticles to move around the system
 (semiclassically, a random walk).
 To avoid such errors, we must make sure that such quasiparticles are not
 created in the first place by performing all braiding operations
 slowly, $t_{\rm op} \gg \Delta$. Again, it is advantageous to make the
 gap $\Delta$ as large as possible.

 While the braiding operations and interference measurements described
 in this paper are similar in the sense that they both involve topological
 operations, they are actually quite different in an important respect.
 The braiding operations by which we envision manipulating our
 qubits are {\it unitary operations}. They involve moving quasiparticles
 around our system over time scales which must be long compared
 to the inverse of the gap and small compared to the inverse of the error
 rate:
 \begin{equation}
 \frac{1}{\Delta} \ll t_{\rm op} \ll T\,e^{\Delta/T}
 \end{equation}
 So long as this order of time scales is respected and
 the quasiparticles are kept far apart (compared to the magnetic length,
 which is the only length scale in the problem), then the system
 is topologically-protected: the quantum state of the system evolves
 precisely as we specify.
 
 On the other hand, our interference measurements are dissipative DC measurements
 (since they require non-zero $\sigma_{xx}$). As far as the multi-qubit
 Hilbert space is concerned, these are not unitary operations but,
 rather, projection onto specified states. It is worthwhile thinking a little more about
 how this `wavefunction collapse' occurs because these measurements are
 potentially noisy. For instance, if a current-carrying edge quasiparticle
 should scatter inelastically, then the interference between its two
 possible trajectories will be spoiled. If this inelastic scattering rate becomes
 too large, then we will be unable to read the state of our quasiparticles
 through an interference measurement because we won't be able to resolve
 two different values of the longitudinal conductivity. We would like to know when
 this will occur. Also, when this occurs, there is
 an interesting quantum measurement theory problem: does the qubit
 wavefunction still `collapse' even when the inelastic scattering rate
 for the test particles is too high to allow us to distinguish the two
 states of the qubit?
 
 To answer these questions, we begin by considering Figure \ref{fig:read}.
 When a single test quasiparticle tunnels between the two edges (without scattering
 inelastically), its wavefunction becomes entangled with the state
 of the qubit: it is in one state, which we will call $|a\rangle$
 when the qubit is in the state $|0\rangle$ and it is in a different
 state, $|b\rangle$, when the qubit is in the state $|1\rangle$.
 If the initial state of the system were $\alpha|0\rangle + \beta|1\rangle$,
 then it is now
 \begin{equation}
 \alpha|0\rangle|a\rangle + \beta|1\rangle|b\rangle
 \end{equation}
 If $|a\rangle$ and $|b\rangle$ were the same, then there
 would be no entanglement with the test quasiparticle at all
 and the coherent superposition of $|0\rangle$ and $|1\rangle$
 is maintained. However, if $|a\rangle$ and $|b\rangle$ were orthogonal,
 then the entanglement between the qubit and the test quasiparticle
 would completely spoil the coherent superposition of $|0\rangle$ and $|1\rangle$
 (unless they can be disentangled later),
 i.e. the qubit wavefunction is `collapsed'.
  
 For small tunneling amplitudes $t_{MN}$ and $t_{PQ}$,
 both $|a\rangle$ and $|b\rangle$ are concentrated on the
 bottom edge in Figure \ref{fig:read} and there is very little difference between these
 two states of the test quasiparticle. We can write
 \begin{equation}
 \langle b|a\rangle = 1 -\delta
 \label{eq:test-overlap}
 \end{equation}
 with $\delta$ small.
 Hence, a single test quasiparticle does not do an effective job
 of `collapsing' the qubit wavefunction. 
 In order to be an effective measurement, we would like
 the qubit to be in the state $|0\rangle$ with probability $1$
 when the test quasiparticle is in the state $|a\rangle$.
 When (\ref{eq:test-overlap}) holds, the qubit is instead in the state
 $\alpha|0\rangle + \beta(1-\delta)|1\rangle$
 when test quasiparticle is in the state $|a\rangle$.
 
 However, if $N$ test quasiparticles
 tunnel, then they all become entangled with the qubit. The combined state of the qubit
 and test quasiparticles is
  \begin{equation}
 \alpha|0\rangle|a,a,\ldots,a\rangle + \beta|1\rangle|b,b,\ldots,b\rangle
 \end{equation}
 Notice that we now have
 \begin{equation}
 \langle b,b,\ldots,b|a,a,\ldots,a\rangle = (1 -\delta)^N
 \end{equation}
 For $N$ sufficiently large, these states are nearly orthogonal.
 Hence, the two states of the qubit cannot be coherently superposed
 unless the qubit is disentangled from the test quasiparticles. This cannot happen once
 the test quasiparticles leave the system at the current lead and thermalize there,
 which is an irreversible process. Thus, we conclude that the
 qubit wavefunction `collapses': when the test quasiparticles are all in
 the state $|a\rangle$, the qubit is in the state $|0\rangle$ with
 probability $1-(1-\delta)^{2N}$, which is $\approx 1$ for $N$
 large.
 
 Now, consider the effect of inelastic scattering on the
 test quasiparticles. Those test quasiparticles which are inelastically scattered
 do not become entangled with the qubit. As far as measuring the qubit is concerned,
 we can forget about them. However, so long as there is a large
 number $N$ of quasiparticles which coherently encircle the qubit without inelastically
 scattering, the qubit wavefunction will collapse, according to the logic above.
 In principle, we can always ensure that this happens simply by waiting long enough,
 but once the inelastic scattering rate becomes of order $\Gamma_{\rm cr} \sim v/L$
 where $L$ is the device size and $v$ the edge velocity, we would have to
 wait an exponentially long time.
 
 The criterion for actually being able to read the value of the qubit
 is a little different, however. It depends on the resolution of our
 ohmmeter. When the inelastic scattering rate is too high, we won't be
 able to resolve that there are actually two different values of the longitudinal conductivity.
 However, depending on how accurately we can measure the longitudinal
 conductivity, this could occur before the bound $\Gamma_{\rm cr}$ is reached.
 Thus, it is possible that the qubit wavefunction might `collapse' by a measurement
 even though we would not be able to read the result.
 
 We should conclude this section with a comment directed to our
 topologically-inclined readers (perhaps 100\% of those who have come so far).
 Many critical details of the interferometer, such as the inelastic scattering length, travel times,
 quasiparticle `delays', etc., are not topological in nature. How is this to be reconciled with
 the fact that in a topological theory all information on the change of state should be encoded
 by the $2+1$-dimensional spacetime history of the medium? The answer is that
 the topology of various tunneling trajectories gives us operators $h_1$, $h_2$, $\ldots$
 evolving the system from initial to final states. However, experimental details
 determine other (dissipative) terms in the evolution equation of the density matrix
 of the system. In the limit that these other terms are small, a pure quantum state will
 remain pure and different $2+1$-dimensional
 space-time histories will contribute coherently to the evolution of this
 quantum state. When they are large, a pure quantum state will evolve
 into a mixed one and the $2+1$-dimensional
 space-time histories will effectively combine to form a super-operator for
 this mixed state density matrix.

\section{Universal Gate Set Using Some Topologically Unprotected Gates}

For all of its remarkable properties, the Pfaffian state suffers from one serious
drawback: the transformations generated by braiding operations are not
sufficient to implement all possible unitary transformations \cite{Freedman02,BK}.
Hence, these operations do not permit universal quantum computation.
However, we don't need to supplement braiding with much in order
to obtain a universal gate set. In this section, we explain a `quick and dirty'
way of doing this.

\begin{figure}[tbh]
\includegraphics[width=3.25in]{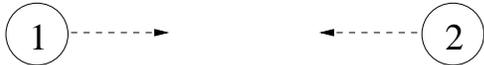}
\caption{By bringing together quasiparticles 1 and 2, which
form a qubit, we can apply the gate $g_1$. This operation is
unprotected and requires control of the distance between
the quasiparticles and the length of time that they are brought
together. However, the required precision might not be very
stringent as a result of the availability of topologically-protected
operations such as $g_3$.}
\label{fig:unprotected}
\end{figure}

First, consider single-qubit operations. If we bring the two quasiparticles
which comprise a qubit close together, as shown in Figure
\ref{fig:unprotected} then their splitting will become
appreciable. This splitting has the form $\Delta\!E(r) \sim e^{- r\Delta/c}$,
where $r$ is the distance
between the quasiparticles and $c$ is some constant with dimensions of
velocity. If we wait a time $T_p$ before pulling the quasiparticles apart again,
then we will apply the phase gate:
\begin{eqnarray}
\label{eqn:phase-unprotected}
{U_P} = \left(
 \begin{array}{cc}
  1 & 0   \\
  0  & e^{i\Delta\! E(r)\,T_p} 
   \end{array}
\right)
\end{eqnarray}
A particularly convenient choice is $\Delta\! E(r)\,T_p = \pi/4$,
which would allow us to apply the same transformation as
the gate $g_1$ described in the next section.

Let us further assume that we can actually perform this operation
on any pair of quasiparticles, not just two quasiparticles
from the same qubit. Note that in order to do this
we only need precise control over the
distance between one pair of
anti-dots since
we can use $g_3$ to move
any desired pair of quasiparticles to these preferred
anti-dots (e.g. using the bucket brigade of
auxiliary anti-dots to move quasiparticles).
If we bring together in precisely the same manner two quasiparticles from different qubits,
then we will couple the two qubits.
In terms of the Majorana modes of (\ref{eqn:braiding-rules}), this
gate is $\exp\!\left(\frac{\pi}{8} {\gamma_i}{\gamma_j}\right)$.
(In the special case that $i$ and $j$ come from the same quasiparticle,
this is the same as (\ref{eqn:phase-unprotected}) up to an overall
phase.)

\begin{figure}[b]
\includegraphics[width=3.25in]{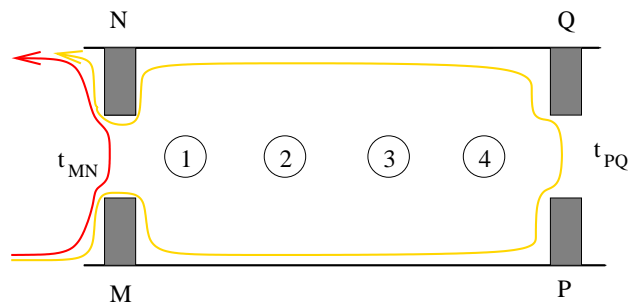}
\caption{A non-demolition measurement of the total topological
charge of two neighboring qubits (comprised of $1$, $2$ and $3$, $4$)
can be done with the interference measurement
shown here. Together with $g_3$ and
the operation $U_P$ shown in Figure \ref{fig:unprotected},
this forms a universal gate set.}
\label{fig:read-4qp}
\end{figure}

The other gate which we need for universal quantum computation
is the non-destructive measurement of the total topological charge
of any pair of qubits. This can be done by using $g_3$ to move one qubit so
that it is next to the other. Then, the interference measurement depicted
in Figure \ref{fig:read-4qp} can
determine the sum of the topological charges of the two neighboring
qubits. This is a measurement of ${\gamma_1}{\gamma_2}{\gamma_3}{\gamma_4}$.
According to ref. \onlinecite{BK-fermions}, this measurement is equivalent
to the application of the gate
$\exp\left(\frac{\pi}{4} {\gamma_1}{\gamma_2}{\gamma_3}{\gamma_4}\right)$
so long as we have the ability to (a) create ancilla in the state $|0\rangle$ and (b)
apply $\exp\left(\frac{\pi}{4} {\gamma_i}{\gamma_j}\right)$, which is simply
the exchange of quasiparticles $i$ and $j$.

An obvious problem is that we have now given up some of the protection
which we have worked so hard to obtain. Even if we could calculate
$\Delta\! E(r)$ with high precision, there would always be some chance
of a mistake in the length of time $T_p$ which the quasiparticles are
close together. Thus, we would be in a situation in which some gates are
exact -- those resulting from braiding operations -- while others are unprotected.
The threshold error rate for the unprotected operations can be much less
stringent (as high as $10\%$), as shown by Bravyi and Kitaev \cite{BK-magic}
for a specific set of perfect gates together
with the noisy creation of a one-qubit ancilla in a specified
state (a `magic state'). It is still an open problem
what the threshold is for the set of protected
gates and the one unprotected gate described above.

\section{Bravyi-Kitaev Construction}
\label{sec:Bravyi-Kitaev}

\begin{figure}[tbh]
\includegraphics[width=2.5in]{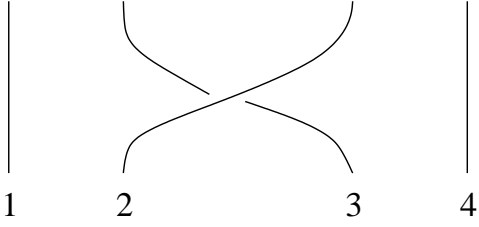}
\caption{Quasiparticles 1 and 2 form a qubit; 3 and 4 form
a second qubit.
Exchanging 2 and 3 applies the gate $g_3$.}
\label{fig:g_3}
\end{figure}

In ref. \onlinecite{BK}, Bravyi and Kitaev constructed a universal set of gates
for a system in the topological phase described by
$SU(2)_2$ Chern-Simons theory, which is the effective field theory \cite{Fradkin98,Fradkin99} 
likely to describe the $\nu=5/2$ quantum Hall state, apart
from an Abelian factor which is unimportant here.
For reference, their gates $\{g_1, g_2, g_3\}$ are:
\begin{eqnarray}
\label{eqn:BK-gate-set}
{g_1} = \left(
 \begin{array}{cc}
  1 & 0   \\
  0  & e^{\pi i/4} 
   \end{array}
\right)\,, {\hskip 0.5 cm}
{g_2} = \left(
 \begin{array}{cccc}
  1 & 0 & 0 &  0 \\
  0 & 1 & 0 &  0   \\
  0 & 0 & 1 & 0 \\
  0 & 0 & 0 & -1
 \end{array}
\right), \nonumber\\
 \textnormal{and } \:{g_3} = \frac{1}{\sqrt{2}}\left(
 \begin{array}{cccc}
  1 & 0 & 0 &  -i \\
  0 & 1 &  -i & 0 \\
  0 & -i & 1 &  0   \\
   -i & 0 & 0 & 1
 \end{array}
\right). {\hskip 1.5 cm}
\end{eqnarray}
$g_1$ is a phase gate on a single qubit. $g_2$ and $g_3$ are
two-qubit gates.
$g_2$ is a controlled phase gate:
if both qubits are in state $|1\rangle$, the state acquires a $(-1)$.
Otherwise, it is unchanged. Alternatively, if we change the basis of the second qubit to
$(|0\rangle\pm|1\rangle)/\sqrt{2}$, then $g_2$ is simply a CNOT gate.
$g_3$ is a two-qubit gate which, together with $g_1$, $g_2$ form
a universal gate set. This particular gate is chosen because it can
be implemented with the simple quasiparticle braiding process
depicted in Figure \ref{fig:g_3}.

The controlled phase gate $g_2$ is more complicated. Suppose we
have two qubits composed of two pairs of quasiparticles
(1,2) and (3,4). We would like
to multiply the state of the system by $(-1)$ only when qubits $(1,2)$
and $(3,4)$ are {\it both} in state $|1\rangle$. The problem is that if 
qubit $(3,4)$ is in state $|1\rangle$ and we take it
around quasiparticle $1$, then a factor of $(-1)$ results,
regardless of whether $(1,2)$ is in the state $|0\rangle$ or $|1\rangle$.
The trick of Bravyi and Kitaev \cite{BK} is to split qubit $(1,2)$ in such a way
as to produce a charge $e/4$ quasiparticle only when $(1,2)$
is in the state $|1\rangle$. If this occurs, then we can take $(3,4)$
around this quasiparticle, and a $(-1)$ will occur if
$(3,4)$ is also in the state $|1\rangle$.

\begin{figure}[tbh]
\includegraphics[width=3.25in]{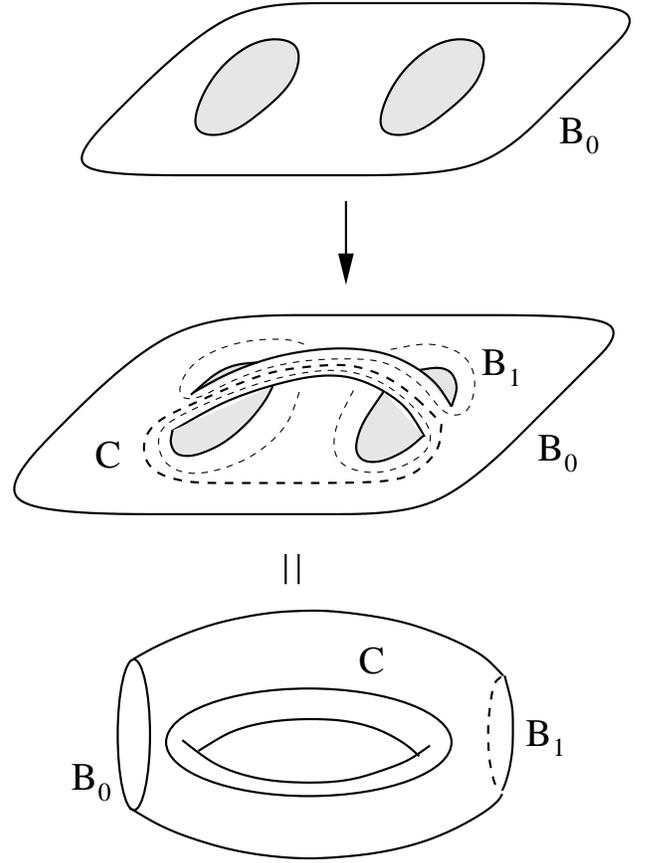}
\caption{By adding an overpass, we connect two antidots.
The dashed curve labeled $B_1$ is the combined boundary of the antidots-plus-overpass. In the implementation of both $g_1$ and $g_2$, we need to check that $W\!({B_1})=1$.
In the case of $g_2$, the controlled
qubit must be taken around $C$ in order to
implement a controlled phase gate on it.
To enact $g_1$, a double Dehn twist must be performed in $C$.
The two anti-dots joined by an overpass in the
middle of the figure are topologically
equivalent to a torus with two punctures
corresponding to $B_0$ and $B_1$ as shown
at the bottom. The meridian of the torus is $C$.}
\label{fig:overpass}
\end{figure}

In order to do this, we perform the following steps which we will
describe here without regard to their feasibility (which will
be taken up in the next section).
Suppose that quasiparticles $1$ and $2$ are
at antidots, which should be understood as punctures in the quantum Hall fluid.
The state of the $(1,2)$ qubit is equal to the topological charge
around the loop $B_0$ in the top diagram in Figure \ref{fig:overpass}.
We will denote this topological charge by $W\!({B_0})$.
We create an overpass which connects these two punctures,
as depicted in the middle part of Figure \ref{fig:overpass}. We check with an interferometry
measurement that the boundary of the antidots-plus-overpass,
labeled $B_1$ in Figure \ref{fig:overpass}, has trivial topological
charge, $W\!({B_1})=1$. If it doesn't, we break the overpass
and rebuild it again until we find $W\!({B_1})=1$. We don't need to repeat this very often since the probability
for $W\!({B_1})=1$ is $1/2$ and the probability for
$W\!({B_1})=\psi$ is also $1/2$ (since the isospin $0$ and $1$ quasiparticles have the same quantum dimension
-- i.e. the same zero-temperature entropy per particle --
in $SU(2)_2$ Chern-Simons theory).
Each time we break the overpass, we
return the qubit to its original state. 
This follows from the general principle
(see appendix \ref{sec:adding-handles})
that adding quantum media is reversible
simply by deleting what was added
(whereas deleting quantum media is generally irreversible).

Once we know that $W\!({B_1})=1$, it
follows for reasons which we discuss below
that if $(1,2)$ is in state $|1\rangle$ (i.e. if $W\!({B_0})=\psi$)
then $W\!({C})=\sigma$ in figure  \ref{fig:overpass}.
If this is the case, then taking $(3,4)$ around the loop $C$
multiplies the state by $(-1)$ if $(3,4)$ is in state $|1\rangle$
and leaves it unchanged if it is in state $|0\rangle$.
On the other hand, if $(1,2)$ is in state $|0\rangle$, then
$W\!({C})=(1-\psi)/\sqrt{2}$ in Figure \ref{fig:overpass}, and taking
$(3,4)$ around $C$ doesn't change the state. Therefore, this sequence of operations applies the gate $g_2$ of
(\ref{eqn:BK-gate-set}).

The Chern-Simons theory calculations which lead to this result are
facilitated by the observation that the topology of two anti-dots
joined by an overpass is a torus with two punctures, corresponding to
$B_0$ and $B_1$, as depicted in the bottom diagram of
Figure \ref{fig:overpass}. The curve $C$ is the meridian of the torus.
Once we have made sure that $W\!({B_1})=1$,
we know that we can fill in this puncture
without changing anything; in other words, the system is equivalent
to a torus with one puncture, $B_0$. Observe
that $W\!({B_0})$ is the
state of the qubit: it is either $1$ or $\psi$. We
would now like to determine $W\!({C})$.
This can be obtained from the $S$-matrix of the theory,
which relates the
topological charge around the meridian, $W\!(M)$,
to the topological charge around the longitude, $W\!(L)$.
If $W\!({B_0})$ is $1$,
the $S$-matrix is,
\begin{eqnarray}
\label{eqn:S-matrix}
S^1_{ij} = 
\left(
 \begin{array}{ccc}
  \frac{1}{2} &  \frac{1}{\sqrt{2}}  &  \frac{1}{2} \\
  \frac{1}{\sqrt{2}} & 0 & - \frac{1}{\sqrt{2}} \\
  \frac{1}{2} &  -\frac{1}{\sqrt{2}}  &  \frac{1}{2}  \\
 \end{array}
 \right)\, , {\hskip 0.3 cm} i,j = 1,\sigma,\psi.
\end{eqnarray}
However, $W\!(L)$
is simply the topological charge on each of the antidots,
which is $\sigma$,
(i.e. $(0,1,0)$ in the matrix notation of (\ref{eqn:S-matrix})). Hence, the
$S$-matrix tells us that $W\!({C})$
is the linear combination $(1-\psi)/\sqrt{2}$.
On the other hand, if $W\!({B_0})=\psi$,
then the vanishing of all $S$-matrix elements
$S^\psi_{ij}$ other than $S^\psi_{\sigma\sigma}=e^{-i\pi/4}$
tells us that $W\!({C})=\sigma$.

We now turn to $g_1$.
The first step in the implementation of gate $g_1$
in (\ref{eqn:BK-gate-set}) is
the same as above:
we take the two anti-dots associated with the qubit under
consideration and join them with an
overpass as in Figure \ref{fig:overpass}.
Again, we check that $W\!({B_1})=1$.
Now, however, we act on this qubit by
performing a double Dehn twist on
the curve $C$. This means that we cut along the curve $C$,
thereby forming two boundaries. We rotate one of them by $4\pi$
relative to the other (i.e. perform two twists) and then glue them back together.
Finally, we remove the band, thereby returning the system to a state
of two $\sigma$ quasiparticles, one on each of the two anti-dots.

A Dehn twist on $C$ has an equivalent effect
as a $2\pi$ rotation of the topological charge associated with $C$.
Since the conformal spins
of $1$, $\sigma$ and $\psi$ are, respectively, $0$, $\frac{1}{16}$, and $\frac{1}{2}$,
the effect of performing two successive Dehn twists is simply the identity ${1^2}=(-1)^2$
if $W\!({C})$ is, respectively, $1$ or $\psi$ and it is
${\left(e^{2\pi i/16}\right)^2}=e^{\pi i/4}$ if the
topological charge is $\sigma$.
Since $W\!({C})$ is perfectly correlated
with the value of the qubit, the effect of this sequence of operations is
the gate $g_1$ in (\ref{eqn:BK-gate-set}).
Note that a single Dehn twist would necessarily
change the charge of the two anti-dots by transferring charge
$e/4$ from one to the other.

\section{From Non-Planar Topology to Time-Dependent Planar Topology}
\label{sec:topology-change}

\begin{figure}[b]
\includegraphics[width=3.25in]{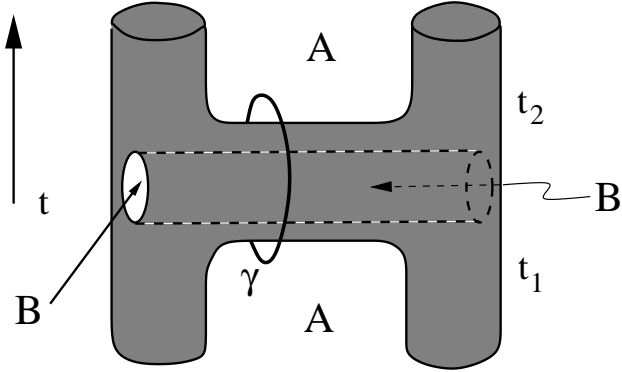}
\caption{A non-planar overpass configuration can be
mimicked by breaking
the region $A$ and adding $B$.
This is done by merging the two anti-dots
between times time $t_1$ and $t_2$. During this merger
interval, we allow the quantum Hall fluid the fill the region $B$.
This faux pass splits the merged anti-dot in the perpendicular direction.
These changes are then undone to return the system to its initial
configuration. In order to verify that removing $A$ was harmless,
we have to perform a tilted interferometry measurement to
check that the topological charge around $\gamma$ is trivial.
See Figure \ref{fig:overpass-time12} for a time slicing of this process.}
\label{fig:faux-overpass}
\end{figure}

The operations described in the
previous section may never be practical in
a real quantum Hall device. Overpasses with high mobility
are implausible, let alone gating them in and out at will.
Dehn twists seem an even more remote possibility.
Fortunately, there are some features of Chern-Simons
theory, which is the effective field theory of our system,
which can be exploited to mimic these types
of operations without leaving the plane or attempting
to perform surgery on our quantum Hall fluid.
In this section, we will explain these features of
Chern-Simons theory and how they can be used
to apply the gates $g_1$ and $g_2$.
Once this problem in topological quantum field theory (TQFT)
has been solved, we turn in the next section to
the new set of problems which arises when we try
to realize this construction in a quantum Hall device.
For the reader who is uninterested in the TQFT details
and wishes to skip ahead to the next section,
we summarize the results of this section: (1) an operation
equivalent to the addition and removal of an overpass between
two antidots can be performed by connecting the two antidots
so long as a curve surrounding the connection (which might be
tilted) has trivial topological charge; (2) a measurement of the topological
charge around a curve $\gamma$ (with a particular framing)
is equivalent to a `Dehn filling' on this curve (which, in turn, is
related to a Dehn twist, as explained below) modulo a few caveats
described in this section. With these two observations
in hand, we can replace `impossible' operations with operations
which are merely very difficult. In section \ref{sec:realizing-BK},
we give concrete illustrations of how this can be done with
the device architecture described in this paper.

\begin{figure}[t]
\includegraphics[width=3.15in]{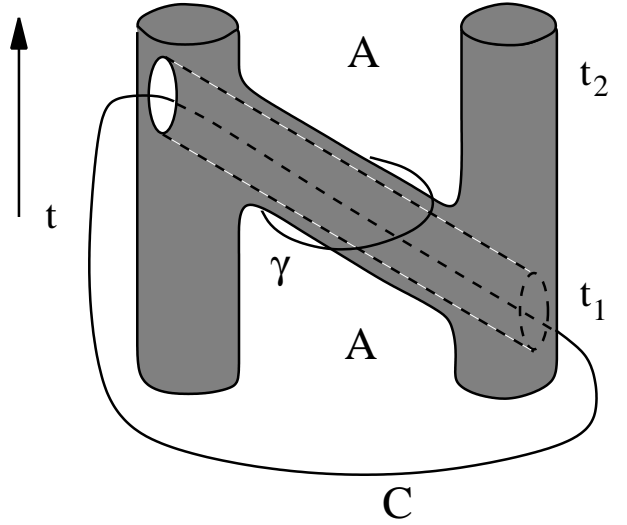}
\caption{Topologically, the process depicted here is equivalent
to that depicted in Figure \ref{fig:faux-overpass}. However,
as a result of the tilt of the faux pass, the triviality
of the charge around the hole in $A$ can be measured with
an untilted measurement. Consequently, the curve $\gamma$ can
now be deformed into a single time slice. The curve labelled
$C$ in this figure is equivalent to the curve $C$ which goes
over the overpass in Figure \ref{fig:overpass}.
This is the curve on which we wish to perform
a Dehn twist. Measuring the topological charge around this curve
allows us to effect the same transformation on the quantum state
of the system without any surgery. See Figure \ref{fig:overpass-tilt1} for a time
slicing of this process.}
\label{fig:faux-overpass2}
\end{figure}

One particularly fortuitous feature of topological field theories,
for our purposes, is the fact that when the topological charge around
a hole is trivial, then the part of the system outside the hole
is impervious to whether the hole is filled in or not.
For example, if the topological charge around $\gamma$
in Figure \ref{fig:spacetime-tilted} is trivial, then this spacetime
history is equivalent, as far as a topological field theory is concerned,
with a spacetime history in which there is no merger whatsoever
between the two anti-dots. This suggests
a way in which we can effectively have overpasses
by taking advantage of the time direction so that `over'
is realized as `at a different time', as shown in Figure \ref{fig:faux-overpass}.
Suppose we want
a band of material $B$ to form an overpass over another band $A$.
We instead break $A$ at time $t_1$, allow $B$ to pass, use $B$ for whatever purpose, break $B$, and then reconstitute $A$ at time $t_2$.
If we could measure the topological
charge around the resulting time-like hole ${t_1}<t<{t_2}$ in $A$ and
if we found that it was trivial, it would be, as far as Chern-Simons theory were concerned, as if $A$ were never broken. The faux
overpass -- or, simply, faux pass -- $B$ is then just as
good as an overpass.

Note that this figure can be continuously deformed to our
convenience. From a topological point of view, Figure
\ref{fig:faux-overpass} is equivalent to Figure
\ref{fig:faux-overpass2}. However, their realizations
are quite different, and they have different practical advantages
and disadvantages. The tilted measurement of $\gamma$
in Figure \ref{fig:faux-overpass} is untilted in Figure \ref{fig:faux-overpass2}.
However, Figure \ref{fig:faux-overpass2} has a moving
anti-dot and island
(whose velocity is the slope of the faux pass in Figure \ref{fig:faux-overpass2}).

\begin{figure}[t]
\includegraphics[width=3.25in]{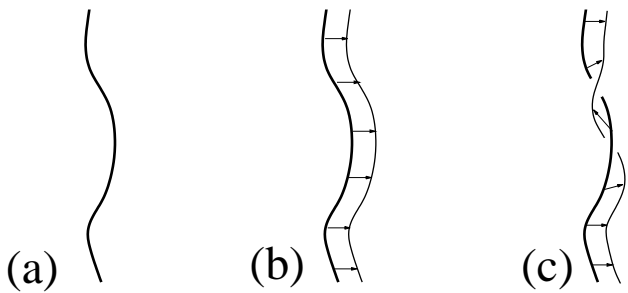}
\caption{The curve in (a) can be framed, for example, as shown
in (b) and (c). The self-linking number in (c) is greater by $+1$ than
in (b).}
\label{fig:framing}
\end{figure}

A second `impossible' operation which we need to perform
is a Dehn twist on a closed curve $C$.
In order to make it possible, we use the following
two facts. (i) It is a fundamental identity of the
``Kirby calculus'' that $(-1)-$framed `Dehn surgery' on a simple linking
circle imparts a $(+1)$ Dehn twist
and, of course, a double Dehn twist arises if
two such surgeries are performed. By `frame', we mean that
the curve $C$ is thickened into a ribbon so that a self-linking
number can be well-defined: it is the linking number of the two
curves formed by the edges of the ribbon (one edge of the ribbon
is $C$, the other edge is traced out by the tip of the
frame vector). Some examples of framed curves are shown
in \ref{fig:framing}. We would like $C$
to be framed so that this self-linking number is $(-1)$.
The meaning of `Dehn surgery' is that a tubular neighborhood (in spacetime) of
the loop $C$ is deleted and then glued back
so that the meridian disk is glued to 
the circle defined by the tip of the frame vector.  Obviously,
physical limitations prevent us from doing this in a quantum Hall device, but 
we can instead (ii) measure the
particle content of a loop $\zeta$ in the interior of a
$2+1-$dimensional space-time. If the result is $1$, we have (up to an
overall normalization factor, corresponding to capping a $2-$sphere)
accomplished Dehn surgery on $C$ as far as Chern-Simons theory
is concerned. This Dehn surgery on a spacetime history has the result
of making it into a history which interpolates between an initial state
before a Dehn twist has been performed on the faux pass
and a final state in which the faux pass has been Dehn twisted.

We would like to explain the relation between
Dehn twist, Dehn surgery, and Dehn filling.
Dehn twist is a method for constructing a diffeomorphism of a surface:
given a closed loop $l$ on a surface, cut the surface along $l$,
twist one side by $2\pi$ and reglue. This is either a $+$ or $-$ Dehn twist
(depending on sign conventions). Our interest in Dehn twist is that it
provides a way to transform a 3-manifold $M$ by cutting it open
along a surface and then regluing using a Dehn twist. To discuss
this procedure, we will simplify matters by
concentrating on an annular neighborhood $A$ of $l$.
(The framing of $l$ defines an annulus $A$, as may be seen
in figure \ref{fig:framing}b,c.) We split the 3-manifold $M$ along $A$,
thereby opening a toridal cavity within $M$. The boundary of this
cavity is a torus $B$ which is just two copies of $A$, joined along their boundaries.
This new manifold with the cavity is essentially $M\backslash\text{neigh}(l)$,
$M$ minus a neighborhood of $l$. The term ``Dehn filling'' refers to
gluing a solid torus back into the cavity; the double process of first
making the cavity and then refilling it is called ``Dehn surgery''.
The possible outcomes of Dehn surgery are parameterized by
the slope of that curve on the cavity boundary $B$ which is matched
to the disk factor $D^2$ of the reglued solid torus ${D^2}\times{S^1}$.
Notice that Dehn twist, from one copy of $A$
to the other, carries a radial arc in one copy to a twisted arc in
the other so that the two mate together to become a
diagonal -- either (1,1) or (1,-1) in the natural $B$-coordinates
(the first component counts the winding number around the ``meridian'' or
shortest direction and the second coordinate counts the winding number
around the ``longitude'' defined by either component of the
boundary of $A$ within $B$) depending on whether the
Dehn twist is $+$ or $-$ respectively. 
Conversely, the instuction to do ``$-1$ Dehn surgery on $l$''
can be expressed as: 1. open the cavity around $l$
(a normal framing on $l$ is required at this point to pick out the surface $A$) 
2. change the correspondence between the two copies of A by a $+1$ Dehn twist.
3. with respect to these new coordinates, Dehn fill the solid torus by
gluing the disk to the meridian of the cavity.
(Steps 2 and 3 together constitute $-1$ Dehn filling since they
tell us to match the $(1,-1)$ curve on the cavity boundary $B$,
which is taken to the meridian by step 2, to the disk within the reglued solid torus.)
Thus cutting and regluing by a $+/-$ Dehn twist is identical to doing $+/-$ Dehn surgery.
The later is simply a more three-dimensional language for the former. Measuring the trivial charge along a curve $C$ is, from the point of view of Chern-Simons theory,
equivalent to supplying a disk (containing no quasiparticles) for the curve $C$ to bound,
or an entire solid torus (with no Wilson loop at its core) for $B$ to bound.
Thus, we propose to accomplish through measurement
a topological operation, namely Dehn twist, which otherwise would have no reasonable
experimental realization.

\begin{figure}[t]
\includegraphics[width=3.15in]{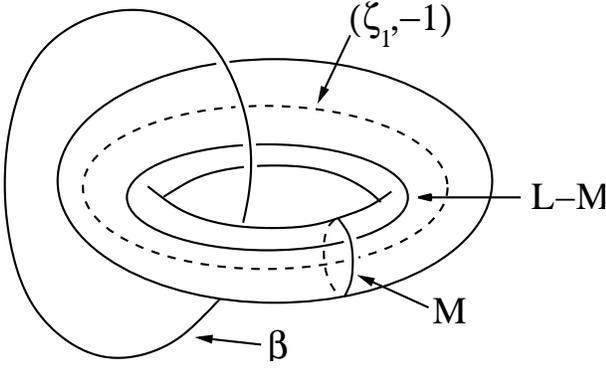}
\caption{If we measure the topological charge around
$({\zeta_1},-1)$ and it is not $1$, then we have performed a Dehn filling on
$({\zeta_1},-1)$ {\it and} left behind a Wilson loop $\beta$, carrying
this topological charge, inside the solid torus.
The state of the system is given by the functional integral
in equation (\ref{eqn:Dehn-state}) for this manifold.
The $3D$ mapping cylinder of a Dehn twist can be described as:
delete a solid torus, then replace it with a twisted Dehn filling.
For convenience, we have drawn the torus in a non-standard way
so that $L-M$ appears simple.}
\label{fig:extra-wilson}
\end{figure}

An important detail to be considered is that measurement might
result in a nontrivial charge. This means, in fact, that the reglued
solid torus does carry a Wilson loop of precisesly that charge.
This Wilson loop is depicted in figure \ref{fig:extra-wilson}
as a loop $\beta$ in the exterior solid torus,
as shown. These states with Wilson loops
correspond to the other two ground states on
the torus (up to an additional factor of $2$ degeneracy
coming from the Abelian part of the theory).
These are the states obtained by performing the Chern-Simons
functional integral over a solid torus with a Wilson loop
carrying topological charge ($\sigma$ or $\psi$ resp.),
but expressed in the meridinal basis.
\begin{equation}
\label{eqn:Dehn-state}
\Psi[a] = \int_{A_{|T^2}=a} D\!A\,\,{W}[A,\beta]\,e^{{\int_M} \cal{L}_{\rm CS}[A]}
\end{equation}
where $\cal{L}_{\rm CS}[A]$ is the Chern-Simons Lagrangian and
$M={D^2}\times {S^1}$ is the solid torus.
$W[A,\beta]$ is the Wilson loop (i.e. the trace of the holonomy
of the gauge field $A$),
$W[A,\beta]={\rm Tr}\left({\cal P}\exp({\oint_\zeta} A)\right)$,
with the trace taken in the fundamental or the adjoint
representation (for, respectively, $\sigma$ or $\psi$),
and the functional integral is over $SU(2)$ gauge fields $A$ such
that $A=a$ on $T^2$, the boundary of the solid torus. Since
the Wilson loop must be around the meridian, $\beta$ is depicted
as `outside' the torus in figure
\ref{fig:extra-wilson}.

In order to perform a double Dehn twist on $C$,
we need to measure the topological charge
around two framed curves which run parallel to $C$.
We will call these curves $(\zeta_1, -1)$ and $(\zeta_2, -1)$;
the $-1$ denotes the framing. Note that
$C$ is equivalent to untwisted copies of the $\zeta$'s,
which can be denoted $(\zeta_1, 0)$ and $(\zeta_2, 0)$.
In order to measure the topological charge
around $(\zeta_1, -1)$ and $(\zeta_2, -1)$,
our measurement quasiparticles will have
to go along a curve with an extra twist;
a way of realizing this is shown in Figure \ref{fig:framed}.
If a measurement finds $W\!({\zeta_1})=1$,
then we have performed the desired
Dehn twist. Finding $W\!({\zeta_1})=\psi$
isn't the end of the world because the extra
Wilson loop which results just gives some extra
minus signs. However, we want to avoid
$W\!({\zeta_1})=\sigma$. One way to do this
is to take a charge $\sigma$ around the loop $(\zeta_1, 0)$.
Then let $({\zeta_1},-1)$ run parallel to the charge,
encircling it to yield the framing $(-1)$.
In Figure \ref{fig:framed}, we have depicted a time-slicing
of such a framed curve: the measurement quasiparticle
must wind around the quasiparticle which is following the
loop $(\zeta_1, 0)$.

We can now use a similar argument to that used
after (\ref{eqn:S-matrix}) to show that
$W\!({\zeta_1})=(1-\psi)/\sqrt{2}$: since
$(\zeta_1 , -1)= L-M$ (i.e. longitude - meridian), we claim that
$W\!({\zeta_1})$ is given by $S T^{-1} |\s\rangle$
where n
\[
{S} =   \left(
 \begin{array}{ccc}
  1/2 & \sqrt{2}/2 & 1/2 \\
  \sqrt{2}/2  & 0 & - \sqrt{2}/2  \\
  1/2 & - \sqrt{2}/2  & 1/2
 \end{array}
\right)
\tn{ and }
{T} =   \left(
 \begin{array}{ccc}
  1 & 0 & 0 \\
  0 & e^{i \pi/8} & 0  \\
  0 & 0  & -1
 \end{array}
\right)
\]
in the $\{1, \s, \psi\}$ basis. 
We check that
$S T^{-1}\,\,|\s\rangle~=~e^{-i\pi/8}\,(|1\rangle~-~|\psi\rangle)/\sqrt{2}$.
Hence, a measurement of $W\!({\zeta_1})$ can only
give $1$ or $\psi$ (with equal probabilities).
By doing this with another charge, parallel to
$\zeta_2$ and encircled by it, we can force
a measurement of $W\!({\zeta_2})$ to be $1$ or $\psi$,
again with equal probabilities, and completely independent
of the result of the $W\!({\zeta_1})$ measurement.

To see that $S T^{-1}$ gives the correct transformation,
note that we wish to transform from the longitudinal basis $L$
to the framing $=k$ basis, $L+kM =$ longitude
$+k$(meridian). To define a basis
$V(T^2)= \sum_{\text{particle types,}a}
V_{a \overline{a}}(S^1 \times I)$,
we need to select a circle, the ``cuff'', to cut the torus along a dual circle, the ``seam'', to trivialize the resulting annulus as $S^1 \times I$.
The transformation is then done in the following two steps\cite{Walker91}:
\begin{multline}
\text{(cuff, seam)}=\\(L,M) 
\stackrel{\textnormal{twist}{^k}}{\longrightarrow}
(L,M+kL) \stackrel{S^{-1}}{\longrightarrow}(M+kL,L).
\end{multline}
Hence, the
composition of these transformations is given by: $S^{-1} T^k S$.

In the discussion above, we have shown that with
two tilted interferometry measurements, we can
accomplish the same thing as a double Dehn twist,
as far as Chern-Simons theory is concerned, with the
caveat that extra Wilson loops might be added to
to the final state.

One important difference between the Bravyi-Kitaev
protocol \cite{BK} and our proposal to mimic this with
time-dependent planar topologies is that, in the latter
case, $W({B_1})$ cannot be measured until the end
of the entire operation. Consequently, we must use the
overpass without knowing if $W({B_1})=1$ or $\psi$
(these are the only two possibilities since two $\sigma$s
can only fuse into these two possibilities). This means that
we must learn to live with the possibility that
$W({B_1})=\psi$. The saving grace is that we will
at least know after the operation whether $W({B_1})=1$ or $\psi$,
which allows us to compensate in the latter case.
Under the assumption that $W({B_1})=1$, we deduced
from the $S-$matrix of the theory that
$W\!({C})$ is perfectly correlated with
$W\!({B_0})$, which is the value of the qubit.
\begin{eqnarray}
W\!({B_0}) = 1 &\Rightarrow& 
W\!({C}) = (1-\psi)/\sqrt{2}
\cr
W\!({B_0}) = \psi &\Rightarrow& W\!({C}) =
\sigma
\end{eqnarray} 
However, the same logic shows that
if $W\!({B_1})=\psi$, then
\begin{eqnarray}
W\!({B_0}) = 1 &\Rightarrow& 
W\!({C}) = \sigma
\cr
W\!({B_0}) = \psi &\Rightarrow& W\!({C}) =
(1-\psi)/\sqrt{2}
\end{eqnarray} 
In other words, the correlation between
$W\!({C})$ and $W\!({B_0})$
has been reversed or, in other words, the roles
of the states $1$ and $\psi$ of the qubit have
been reversed.

In the case of $g_2$, this
applies to the control qubit, on which all of the
operations are performed (the controlled qubit is
passive). Hence, if we find at the end of our procedure
that $W\!({B_1})=1$, then we know that
we have applied the desired phase gate $g_2$.
If , instead, $W\!({B_1})=\psi$, then the gate has inadvertently
interchanged the roles of $1$ and $\psi$
within the controlling qubit so that
\begin{eqnarray}
 {\tilde g_2} = \left(
 \begin{array}{cccc}
  1 & 0 & 0 & 0 \\
  0 & -1 & 0 & 0 \\
  0 & 0 & 1 & 0 \\
  0 & 0 & 0 & 1
 \end{array}
\right)
\end{eqnarray}
has instead been applied.  This is not too serious since
repeated application of the protocol gives a random walk in the
group $Z_2 \bigoplus Z_2 = \{1, {g_2}, {\tilde g_2}, {g_2}{\tilde g_2}\}$. 
Our $W\!({B_1})$ measurements tell us where we are within $Z_2
\bigoplus Z_2$ as we randomly walk; we simply halt upon reaching
$g_2$. The tails on a ``long walk'' decay exponentially so
this delay is acceptable.

Now consider $g_1$. Let us first suppose, for simplicity,
that $W\!(\zeta_1) = W\!(\zeta_2)=1$.
If we find at the end of our procedure
that $W\!({B_1})=1$, then we know that
we have applied
$ {g_1} = \left(
 \begin{smallmatrix}
  1 & 0 \\
  0 & e^{\pi i/4}  
 \end{smallmatrix}
\right) $.
However, if we find that $W\!({B_1})=\psi$,
then we have applied
\begin{eqnarray}
\label{eqn:g_1^-1}
\left(
\begin{array}{cc}
   e^{\pi i/4} & 0  \\
               0 & 1
 \end{array}
\right) = e^{\pi i/4} \: g_1^{-1} .
\end{eqnarray}
There is one added complication in the case
of $g_1$, as compared to $g_2$: we have the
added uncertainty in the outcome of the
$W\!(\zeta_1)$, $W\!(\zeta_2)$ measurements.
If $W\!(\zeta_1) = W\!(\zeta_2)=\psi$,
then two Wilson loops carrying $\psi$
appear within the solid torus. However,
these two Wilson loops fuse to form $1$,
which is again trivial, just as in the case
$W\!(\zeta_1) = W\!(\zeta_2)=1$. However, if 
$W\!(\zeta_1)\cdot W\!(\zeta_2) = \psi$,
which is as good as a single Wilson loop $\zeta$
carrying topological charge $\psi$ parallel to $C$,
there can be a non-trivial effect.
If $W\!({C})=(1-\psi)/\sqrt{2}$,
which means that $W\!({B_0})\cdot W\!({B_1})=1$,
the Wilson loop contributes no extra phase.
However, if $W\!({C})=\s$, which means that
$W\!({B_0})\cdot W\!({B_1})=\psi$,
then the Wilson loop $\zeta$ encircles the
same topological charge $\sigma$ which is
encircled by $C$ (to which it is parallel).
When a $\psi$ encircles a $\sigma$, a $(-1)$
results. This means that if $W\!(\zeta_1)\cdot W\!(\zeta_2) = \psi$,
then instead of $g_1$ or $g_1^{-1}$, we have acutally applied
${\sigma_z}\,{g_1}=g_1^{-3}$ or ${\sigma_z}\,g_1^{-1}=g_1^{3}$,
where ${\sigma_z}=\left(
 \begin{smallmatrix}
  1 & 0 \\
  0 & -1  
 \end{smallmatrix}
\right)$.

In all eight measurement outcomes
for $W\!(\zeta_1)$, $W\!(\zeta_2)$
we have, up to an overall phase,
implemented either $g_1$, $g_1^{-1}$, $g_1^{-3}$, or $g_1^{3}$.
Thus our protocol generates a type of random walk on $Z_8$.
Since we know the measurement outcomes we may iterate
the protocol until we arrive at $g_1$,
which is again efficient.

Therefore, we conclude that finding $W\!({B_1})=\psi$
is not a calamity for the implementation of
either $g_1$ or $g_2$.

\section{Realizing the Bravyi-Kitaev Gates with Time-Dependent
Planar Topology and Tilted Interferometry}
\label{sec:realizing-BK}

In this section, we will discuss how the operations
of the previous section, which are merely difficult,
rather than impossible, might be implemented in a
quantum Hall device. We need to be able to do four
things: (1) move our qubits at will, (2) create `faux
passes' -- which are equivalent to overpasses in
Chern-Simons theory -- with the
spacetimes histories depicted
in Figure \ref{fig:faux-overpass} or  Figure \ref{fig:faux-overpass2},
(3) measure the topological charge around a tilted trajectory
such as $C$ in Figure \ref{fig:faux-overpass2}, and (4)
create a quasiparticle pair and move them around as desired before
annihilating them. We have already discussed (1) and (3) in
section \ref{sec:braiding-interfero}. We now turn to (2) and (4).

\begin{figure*}[t]
\includegraphics[height=5in]{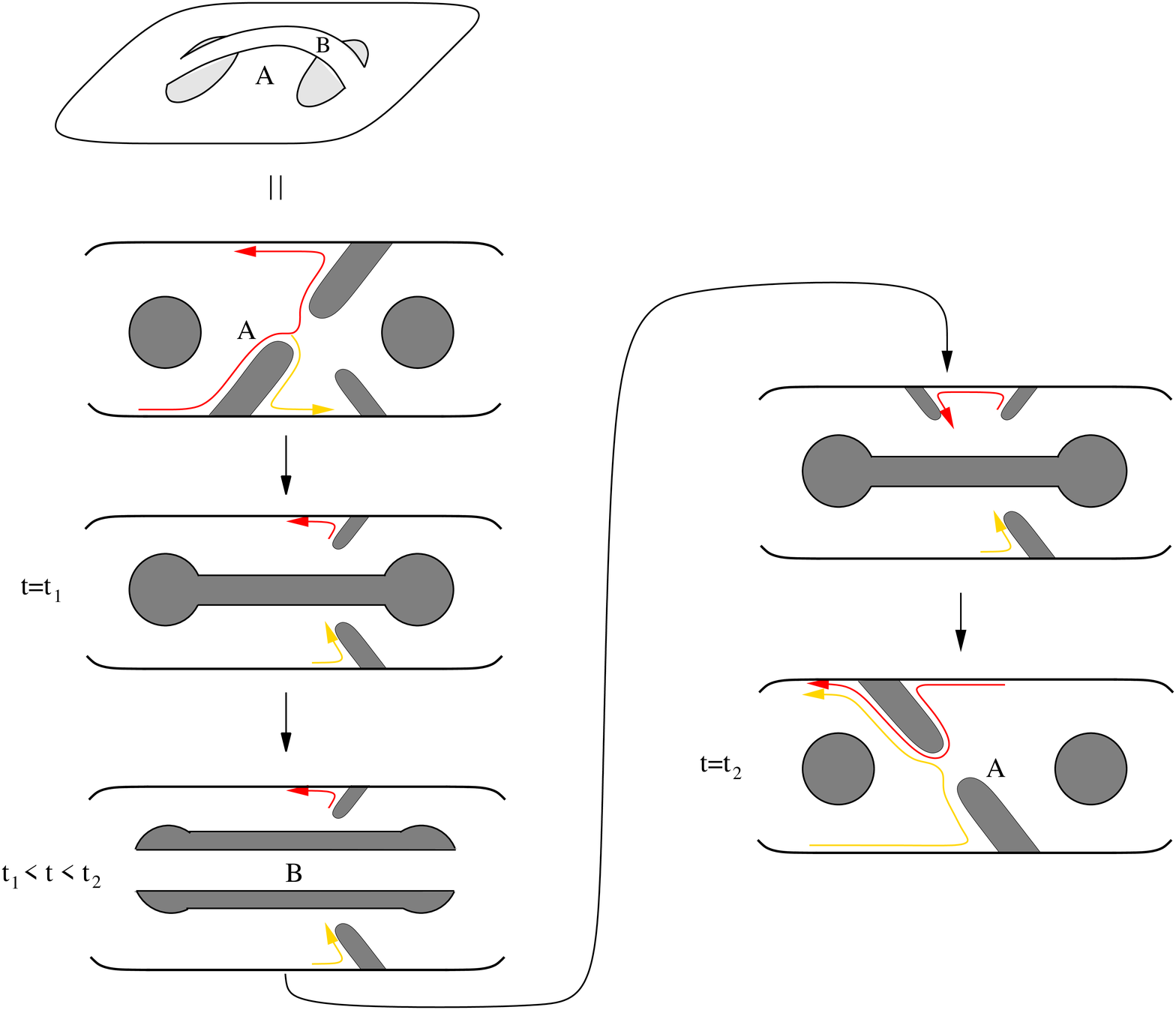}
\caption{A non-planar overpass configuration can be mimicked by
breaking the region $A$ and adding $B$.
This is done by merging the two anti-dots at time $t_1$ and splitting them in the other direction. The faux pass $B$ is then used to either
apply $g_1$ or $g_2$, as described in Section \ref{sec:topology-change}.
These changes are then undone to return the system to its initial
configuration. In order to verify that removing $A$ was harmless,
we have to perform a tilted interferometry measurement in which
the red and yellow trajectories interfere.}
\label{fig:overpass-time12}
\end{figure*}

The spacetime history shown in Figure \ref{fig:faux-overpass}
can be realized by the sequence of steps depicted in
Figure \ref{fig:overpass-time12}.
In this figure, the region $A$ (the `underpass') of
the quantum Hall fluid separates the two anti-dots.
At time $t_1$, region $A$ is broken so that the two anti-dots
are joined into one large oblong anti-dot.
After this occurs, a strip of quantum Hall fluid is allowed
to split the large anti-dot in the perpendicular direction
(the bottom left picture in Figure \ref{fig:overpass-time12}).
This is the faux pass $B$ which plays the role of
the overpass.  The spacetime
region carved out by this strip is the tube in
Figure \ref{fig:faux-overpass}.

In order to check that the
topological charge around the time-like hole is trivial,
we need to do a tilted interferometry measurement similar to that
depicted in figures \ref{fig:spacetime-tilted}, \ref{fig:tilted-traj}.
The interference between the red and yellow curves in Figure
\ref{fig:overpass-time12} measures the topological charge
around the hole in $A$. There is an
obvious drawback here, which is that the result of this
measurement will not be know until after the entire
procedure is complete. We will return to this issue later.
For now, let us consider the other operations which we
need to perform.

\begin{figure}[t]
\includegraphics[width=3in]{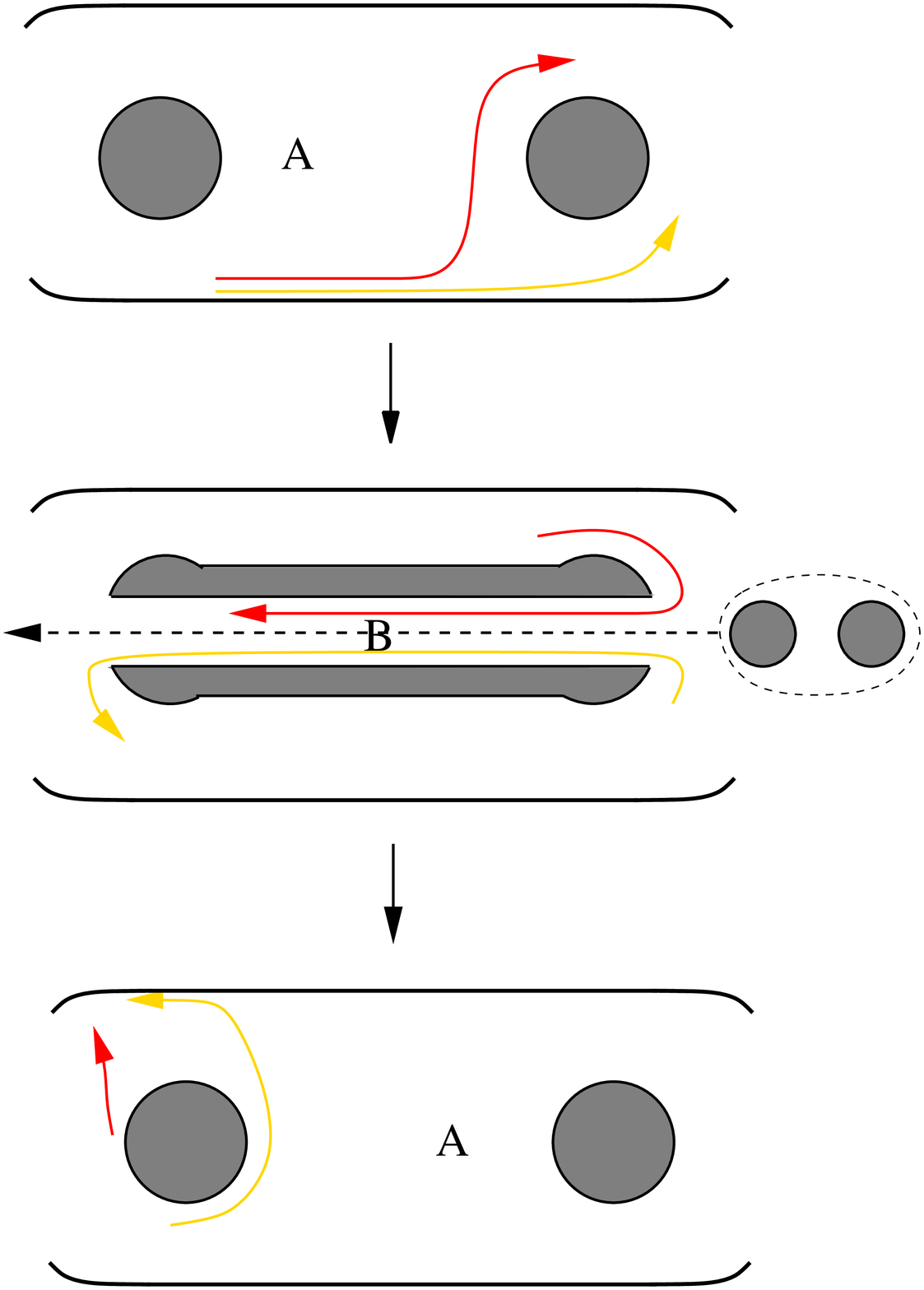}
\caption{The interference between the red and yellow
trajectories is a measurement of the topological charge around
$B_1$. In order to apply $g_2$, we take the controlled qubit
across the faux pass}
\label{fig:overpass-B1}
\end{figure}

\begin{figure}[t]
\includegraphics[width=3.25in]{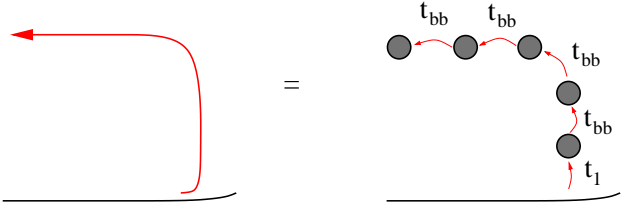}
\caption{The quasiparticle trajectories shown in Figures
\ref{fig:overpass-time12} and \ref{fig:overpass-B1}.
may actually look more like the trajectory on the right above:
a series of hops from one small anti-dot to another.}
\label{fig:bucket-brig}
\end{figure}

During the time that the faux pass region $B$ is available,
we must take our qubit over it and then check whether
the topological charge around the curve $B_1$ is 
$1$ or $\psi$. These are both depicted in
Figure \ref{fig:overpass-B1}.
Both of these processes must
occur while those depicted in Figure \ref{fig:overpass-time12}
are simultaneously occurring.
We envision doing both of these with a bucket brigade
of anti-dots which are used to ferry both the measuring test quasiparticles
and the controlled qubit across the faux pass, as shown in
Figure \ref{fig:bucket-brig}.
This is relatively straightforward for the qubit: the two quasiparticles
comprising the qubit must be moved across the region $B$
in figure \ref{fig:overpass-time12}.
Of course, this process must occur without the two quasiparticles
of the controlled qubit fusing. Hence, we need to keep them
far apart, either with a large faux pass or by having one follow at a
distance behind the other.

The test quasiparticle with which
we measure $B_1$ is trickier. Consider the
red and yellow trajectories in Figure \ref{fig:overpass-B1}.
There should be some small amplitude $t_r$ for a quasiparticle
at the bottom edge to tunnel to the top edge via
the red trajectory, which takes it over the faux pass,
and a small amplitude $t_y$ for it
to tunnel via the yellow trajectory. In the `bucket brigade'
scenario, the red trajectory is actually composed of a series
of hops from one anti-dot to another. Let us assume that
the amplitude for a quasiparticle to hop from the edge
onto the first anti-dot is $t_1$. Let us further assume
that the amplitude for it to hop from the first anti-dot to
the second anti-dot is $t_{bb}$; from the second to the
third, again $t_{bb}$; and so on until the quasiparticle
finally hops onto the top edge. Then
${t_r} \sim {t_1}\,t_{bb}\,t_{bb}\ldots t_{bb}$, where the number
of factors of $t_{bb}$ depends on how many anti-dots are
in the bucket brigade. We need this amplitude to be small
so that the topological order of the state is not degraded,
but is should be large enough to be measurable.
This might be most easily done by making $t_1$ small
and $t_{bb}$ not too small. Of
course, the same reasoning holds for the yellow trajectory.
Finally, we need the amplitudes for these two processes to interfere coherently. This means that the coherence time for a quasiparticle
on either one of these trajectories must be longer than the
time of flight, which might be difficult to ensure.

\begin{figure}[t]
\includegraphics[width=2.5in]{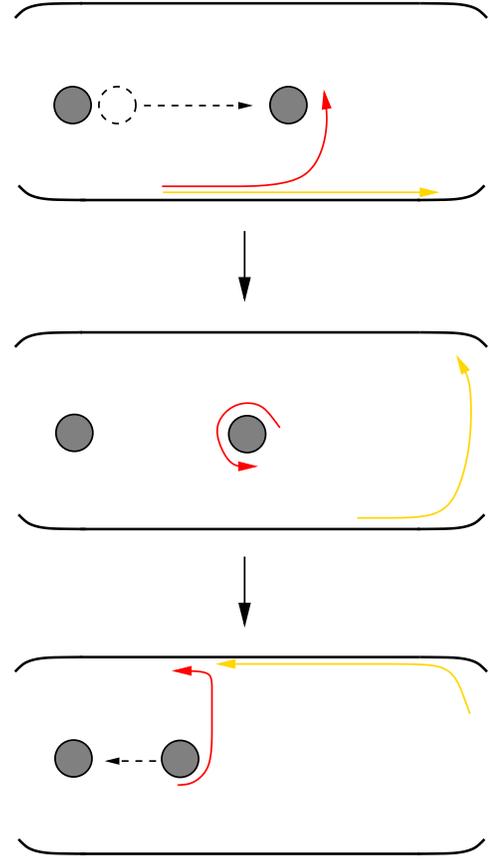}
\caption{In order to apply $g_1$, we must
create a pair of auxiliary quasiparticles, take one over the
faux pass (not shown), and then annihilate them again. While
this is occurring, we must measure the topological
charge around a curve which follows the auxiliary
quasiparticle over the faux pass, while encircling
it. By encircling the auxiliary quasiparticle (presumably while
hopping along a bucket brigade of anti-dots), this trajectory
obtains a non-trivial framing: it is the framed curve $({\xi_i},-1)$.
A measurement of
the interference between the red and yellow curves
can determine the topological charge around this
(tilted) framed curve.}
\label{fig:framed}
\end{figure}

In order to implement $g_1$, we need to also send
an auxiliary quasiparticle over the faux pass. This can be
done as shown in Figure \ref{fig:framed}, perhaps using
a `bucket brigade', as in Figure \ref{fig:bucket-brig}.
We must then measure the topological charge
around the curve $({\zeta_1},-1)$,
which means that we must measure the interference
between the trajectories in Figure \ref{fig:framed}. We do this
with $({\zeta_2},-1)$ in the same way.

\begin{figure*}[t!]
\includegraphics[height=6.5in]{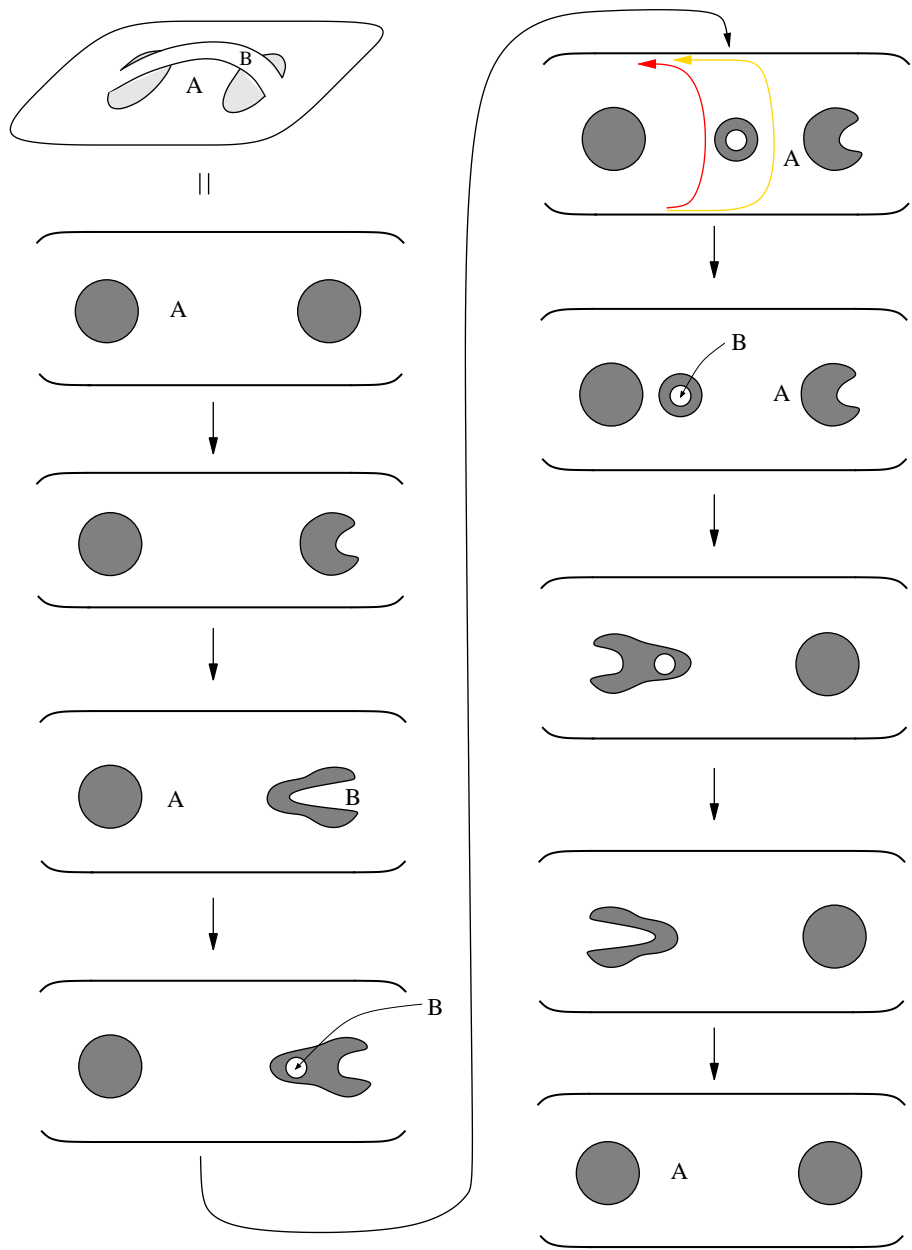}
\caption{By sending an annular intermediary between the
two anti-dots, we can construct the spacetime history
of Figure \ref{fig:faux-overpass2}. The shuttle between
the two antidots is annular and the island of quantum
Hall fluid in the middle of the annulus is the faux pass $B$.
A measurement of the interference between the red
and yellow trajectories can check that the tilted hole
in region $A$ is topologically trivial as far as
Chern-Simons theory is concerned.}
\label{fig:overpass-tilt1}
\end{figure*}

There is a problem with the procedure described
in these figures, which is that the configuration shown in
the third picture in Figure \ref{fig:overpass-time12}
has two anti-dots merged. Consequently, the state of the
two anti-dots does not have exponential protection. The splitting
between the two states of the qubit formed by this pair of anti-dots
is $w\sim 1/x$, where $x$ is the linear extent of the merged antidot.
This can actually be avoided in the following way. Instead of
merging the two anti-dots, we send an intermediary
which shuttles from one to the other. This is done by
breaking the anti-dot on the right into two anti-dots,
one with electrical charge $e/4$ and the other
with no electrical charge.
We then move the neutral anti-dot to the left and
merge it with the left anti-dot. This would simply replace the merger by
the `tilted merger' shown in Figure \ref{fig:faux-overpass2}.
In order to have an overpass, we need the anti-dot to
be annular so that there is a region of $\nu=5/2$ quantum Hall fluid
in the middle of it, as shown in Figure
\ref{fig:overpass-tilt1}. While this moving shuttle
is between the two anti-dots, we
check by ordinary interferometry that it carries topological charge $1$,
rather than $\psi$.
If it doesn't, then we re-merge it with the
right anti-dot and repeat the same process until we find that
the topological charge around the shuttle is $1$. The abortive attempts at this do not affect the qubit (except
possibly by an irrelevant overall phase): splitting $a$ into $b
\otimes c$ and then re-fusing results in the original particle type,
so it is a multiple of the identity.  The qubit is
clearly unaffected since the phase resulting from
such operations is independent, by locality,
of the state of the qubit.
In order to apply $g_2$, the controlled qubit
must sit in the interior of the shuttle as it moves from
one anti-dot of the control qubit to the other.
In order to apply $g_1$, the auxiliary quasiparticles
must do so. It must be noted that
this approach again has the difficulty that the coherence times
of complicated interfering quasiparticle trajectories must be kept long.

\section{Discussion}

In this paper, we have discussed how the $\nu=5/2$ quantized Hall plateau
can be used as the basis of a quantum computer, assuming that this plateau
is in the universality class of the Pfaffian state. Pairs of charge $e/4$
quasiparticles form qubits. We propose to pin the quasiparticles
at anti-dots so that by moving the anti-dots, we move the quasiparticles.
The two states of a pair of quasiparticles, $|0\rangle$ and $|1\rangle$, can be identified
with the primary fields $1$, $\psi$ of the conformal field theory of
the critical $2D$ Ising model, or with the spin $0$ and spin $1$
representations of $SU(2)_2$ Kac-Moody algebra
(or the related quantum group). The value of any qubit can be read by a
transport measurement which is sensitive to the interference between
two possible quasiparticle trajectories encircling the qubit. However, local measurements
cannot distinguish the two states of the qubit so long as the two
quasiparticles are kept apart. The error rate is
astronomically low, so these qubits form
an essentially perfect quantum memory \cite{DasSarma05}.
Two simple gates can
be implemented by quasiparticle braiding:
(1) tunneling a quasiparticle from the edge between
the two quasiparticles comprising a qubit and (2) by exchanging a
quasiparticle from one qubit with a quasiparticle from another qubit.

In order to be able to apply any possible unitary transformation to
our qubits -- i.e. in order to have universal quantum computation --
we can try either of two approaches. One is to use the unprotected operation
of bringing together the two quasiparticles comprising a qubit so
that a phase gate will result from the resulting energy splitting
between their two states. Although this operation is unprotected,
the error threshhold may not be too strict because the other
operations are exact due to topological protection. 
The other approach is to use some complicated manipulations
of the anti-dots which involves moving, splitting, and rejoining them
in order to fool the topological field theory governing the Pfaffian
state into behaving as if we have performed non-embeddable topology changes. Although complicated, these operations allow, in principle, for
a universal set of exact gates. By now the reader may have decided
that the former approach is more promising in the short run because
it does not require extremely complicated quasiparticle manipulations.
However,  if technological advances make the latter approach more
feasible, then it has the virtue of complete topological protection.
Also, the beautiful topology involved lends it an intrinsic `coolness'
factor.

If the weaker $\nu=12/5$ plateau proves to be a non-Abelian state related
to $SU(2)_3$ Chern-Simons theory, then neither of these two
compromises would be necessary. We would again realize qubits as above,
but, in this case, the braid group alone would be sufficient
for universal quantum computation.

\acknowledgements
We would like to thank S. Das Sarma,
M.~P.~A. Fisher, A. Kitaev, C. Marcus, K. Shtengel,
J. Slingerland, and A. Stern for discussions.
We would especially like to thank
P. Bonderson for discussions and a careful reading
of earlier versions of this paper.
We would also like to acknowledge the
hospitality of the Kavli Institute for Theoretical Physics at UCSB
and the Aspen Center for Physics.
We have been supported by the ARO under Grant
No.~W911NF-04-1-0236. 
C.~N. has also been supported by the NSF
under Grant No.~DMR-0411800.

\appendix

\section{Effective Field Theory approach to the Pfaffian State}
\label{sec:EFT}

The multi-quasihole structure which we encountered in section
\ref{sec:qubit}  is precisely the same structure as occurs in the Ising model.
Indeed, the motivation of Moore and Read for first writing down
the Pfaffian wavefunction was its relation to conformal blocks
of the Ising model\cite{Moore91}.
The first hint that there might be such a connection comes from the observation
that the Pfaffian factor in the wavefunction
(\ref{grdstate}) is equal to a correlator of $\psi$ fields:
\begin{equation}
\left\langle \psi({z_1}) \, \psi({z_2})\,\ldots\,\psi({z_{2n}})\right\rangle
= {\rm Pf}\!\left( \frac{1}{z_j - z_k }\right)
\end{equation}
Multi-quasihole wavefunctions are related to conformal blocks
with spin fields such as \cite{Moore91,Nayak96c}. Before
discussing this in any more detail, however, we should first explain
why the quasiparticles of the Pfaffian state have anything to do with
the primary fields of the Ising model.

The answer lies in the connection discovered
by Witten \cite{Witten89} between $2+1$-D Chern-Simons theories and
$1+1$- or $2$-D conformal field theories (CFTs). According to Witten \cite{Witten89}
(see also the explicit constructions of Elitzur {\it et al.} \cite{Elitzur89}), the Hilbert space
of states of a Chern-Simons theory on a $3D$ manifold $M=X\times R$
is equivalent to the vector space of conformal blocks of an associated 
CFT on the $2D$ manifold $X$. The Hilbert space of a Chern-Simons
theory with sources located at ${\bf r_1}, {\bf r_2}, \ldots, {\bf r_n}$ carrying
representations ${\rho_1}, {\rho_2}, \ldots, {\rho_n}$ is equivalent
to the conformal blocks of an $n$-point correlation function of primary fields
associated with these representations.

If we have an electron system
which is described at long-wavelengths by a Chern-Simons theory,
we can use the wavefunctions of the Chern-Simons theory with sources
as approximate wavefunctions of the electron system. By
the above connection, this means we can use conformal blocks in the associated CFT
as wavefunctions of the electron system. In each of these two
steps, the essential features which are preserved are the braiding properties
of the quasiparticles of the system. The different primary fields
in the CFT correspond to the different topological charges in
the Chern-Simons theory, which, in turn, correspond to the different
quasiparticle species in the electron system.

As shown in ref. \onlinecite{Fradkin98,Fradkin99}, the low-energy
effective field theory associated with the Pfaffian state of bosons
at $\nu=1$ is an $SU(2)_2$ Chern-Simons theory. The effective
field theory for fermions at $\nu=1/2$ (or other even-denominator
filling fractions and for bosons at other odd-denominator filling fractions)
is a deformation of this theory: a Higgs-ed $SU(2)_2$ Chern-Simons theory
(independent of the filling fraction)
coupled to a $U(1)$ Chern-Simons theory
with a filling-fraction-dependent coupling constant.
The CFT corresponding to this Chern-Simons theory is
the tensor product of an $SU(2)/U(1)$ coset model
with a $c=1$ $U(1)$ factor \cite{Moore91}. However,
the $SU(2)/U(1)$ coset model is equivalent to the Ising model.
According to this correspondence, the primary fields which
are associated with half-flux quantum quasiparticles are
Ising spin fields (multiplied by a field from the $c=1$ sector).

The $U(1)$ factor, which accounts for the charge, gives the
correct Abelian factors for quasiparticle statistics and contributes
a factor to the ground state degeneracy. The more non-trivial physics
is contained in the $SU(2)$ part of the theory. Hence, we can be
a little sloppy and drop their $U(1)$ quantum numbers and
refer to quasiparticles by their $SU(2)_2$ quantum numbers
(which we will call `isospin'). According to this nomenclature,
there are three primary fields, with isospins $0,\frac{1}{2},1$,
corresponding to $1,\sigma,\psi$ of the Ising model. These
are, respectively, the vacuum, the half-flux quantum quasiparticle,
and the neutral fermion. The topological degeneracy of multi-quasiparticle
states reflects the decomposition of the product of two
isospin $1/2$s: $\frac{1}{2}\otimes\frac{1}{2}=0\oplus 1$.
Thus, there are two different reasons why it is
natural to call the two states of a qubit $0$ and $1$.

Returning now to the conformal blocks which model
wavefunctions of electrons at $\nu=1/2$ (in the first excited Landau level),
we note that the corresponding CFT contains both the $c=1/2$ Ising model
and a $c=1$ chiral boson (accounting for the electrical charge):
\begin{equation}
{\cal S} = \int {d^2} z \, \left(\psi \,{\partial_{\overline z}} \psi + 
({\partial_z} \phi) ({\partial_{\overline z} \phi})\right)
\end{equation}
We retain only the right-handed part $\varphi$ of $\phi = \varphi + {\overline \varphi}$,
with ${\partial_{\overline z}}\varphi = {\partial_z}{\overline \varphi} = 0$.
If we assume that the operator corresponding to the electron is:
\begin{equation}
\Psi_{\rm el}(z) = \psi(z)\,e^{i\varphi(z)\sqrt{2}} 
\label{eqn:CFT-electron-op}
\end{equation}
then we have
\begin{multline}
\left\langle \Psi_{\rm el}({z_1})\,\Psi_{\rm el}({z_2})\,\ldots\,\Psi_{\rm el}({z_{2N}})
\,e^{\int {d^2}z \rho\,\varphi(z)\sqrt{2}}\right\rangle
=\\ \prod_{j<k} (z_j - z_k)^2 \prod_j e^{- |z_j|^2/4 }
  \cdot {\rm Pf}\!\left( \frac{1}{z_j - z_k }\right)~.
\end{multline}
The last term in the correlation function on the left-hand-side of this
equation is a neutralizing charge background without which
the Coulomb gas correlation function would vanish.
To find the primary fields of this $c=3/2$ theory, we need to
put together the primary fields of the (right-handed chiral
component of the) Ising model -- the identity, $1$; the Ising spin field, $\sigma$;
and the Majorana fermion, $\psi$ -- with exponentials of the chiral
boson $\varphi$.  The primary fields must be local with respect to the
electron operator -- i.e. quasiparticles can have non-trivial statistics with
each other but they must have trivial statistics with respect to an electron
since an electron wavefunction must be single-valued in electron coordinates.
A field which satisfies this condition is
\begin{equation}
\Phi_{\rm qp}(\eta) = \sigma(\eta)\,e^{i\varphi(\eta)/2\sqrt{2}} 
\end{equation}
This field corresponds to the charge $e/4$ quasiparticle:
\begin{multline}
\biggl\langle \Phi_{\rm qp}({\eta_1})\Phi_{\rm qp}({\eta_2})\,\times\\
\Psi_{\rm el}({z_1})\Psi_{\rm el}({z_2})\ldots\Psi_{\rm el}({z_{2N}})
\,e^{\int {d^2}z \rho\,\varphi(z)\sqrt{2}}\biggr\rangle
=\\ 
   {\rm Pf}\!\left(
  \frac{\left({z_j}-{\eta_1}\right)\left({z_k}-{\eta_2}\right)+{z_j}\leftrightarrow{z_k}}{z_j - z_k}\right)
  \,\times\\
  \prod_{j<k} (z_j - z_k)^2 \prod_j e^{- |z_j|^2/4 }
\end{multline}
The four-quasihole wavefunctions in (\ref{eqn:fourqh}) are
given by the conformal blocks of $\langle \Phi_{\rm qp}\Phi_{\rm qp}\Phi_{\rm qp}\Phi_{\rm qp}
\Psi_{\rm el}\Psi_{\rm el}\ldots\Psi_{\rm el}\,\exp({\int {d^2}z \rho\,\varphi\sqrt{2}})\rangle$.
The two different wavefunctions correspond to the two different fusion channels
$\sigma\cdot\sigma \sim 1 + \psi$. (Furthermore, the two conformal blocks are
the wavefunctions (\ref{eqn:fourqh}) in the special basis in which the explicit
monodromy of the wavefunction gives the complete braiding
matrices \cite{Gurarie97}).

\begin{figure}[tbh]
\includegraphics[width=3.45in]{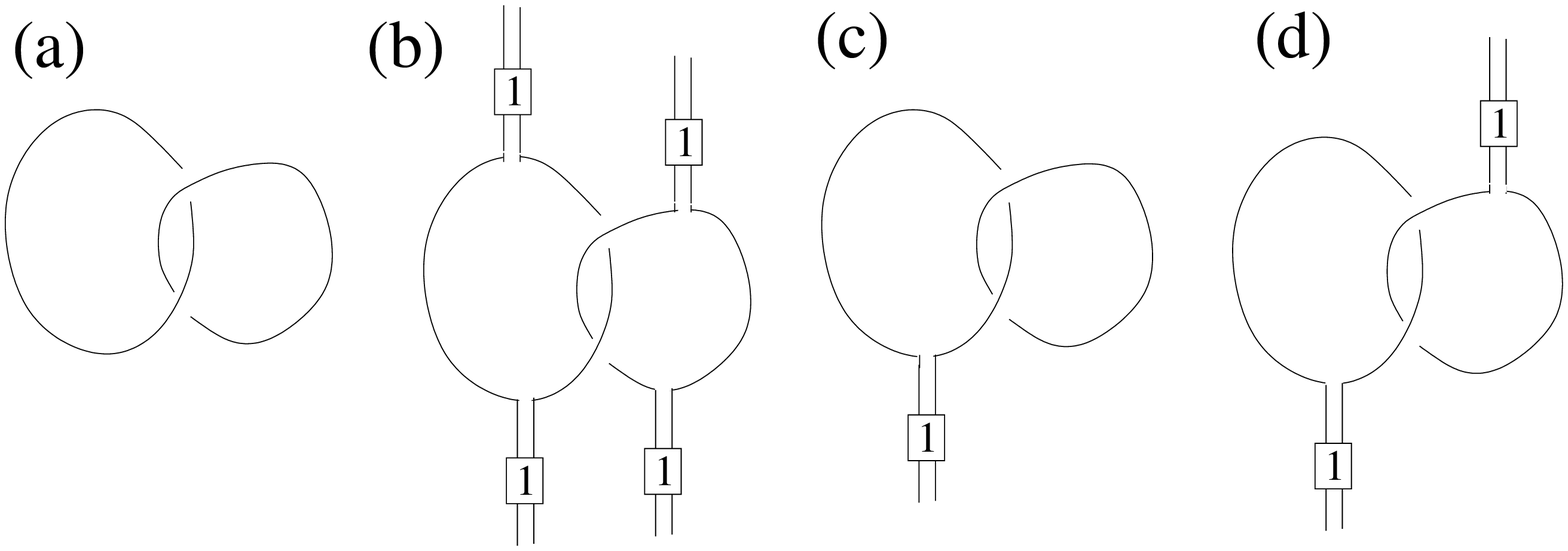}
\caption{By evaluating the Jones polynomial at $q=\exp(\pi i/4)$ for these links,
we can obtain the desired matrix elements for braiding operations manipulating the qubit.
The boxed $1$ is a projector on the pair of quasiparticles which puts them in the
state $|1\rangle$.}
\label{fig:Jones-qubit}
\end{figure}

The full set of primary fields is given in the table below.
The columns correspond to Ising model primary fields while the rows correspond to
the $c=1$ primary fields. The entries of the
table are the quasiparticle electrical charges. Each quasiparticle
corresponds to an operator formed by multiplying the operator
at the top of its column by the operator to the left of its row.
Note that it is not a naive tensor product
between the two theories because the quasiparticles containing $\sigma$
are offset  by $e^{i\varphi/2\sqrt{2}}$ from those containing $1,\psi$.
(Again, these assignments are determined by the requirement of locality
with respect to the electron operator (\ref{eqn:CFT-electron-op}).)
{\vskip 0.25 cm}
\centerline{\begin{tabular}{c|c|c|c|}
             & {\hskip 0.3 cm} $1$   {\hskip 0.3 cm}  &
             {\hskip 0.25 cm}$\sigma\,e^{i\varphi/2\sqrt{2}}$  {\hskip 0.25 cm}   &
             {\hskip 0.3 cm} $\psi$ {\hskip 0.3 cm} \\ \hline
             & & &      \\
$1$      & 0    & $e/4$   & 0\\ 
             & & & \\  \hline
             & & &      \\
 $e^{i\varphi/\sqrt{2}} $ & $e/2$             & $3e/4$    & $e/2$\\ 
 & & &      \\
 \cline{2-4}
\end{tabular}}
{\vskip 0.5 cm}
One of the nice features of the relationship between the Pfaffian state
and $SU(2)_2$ Chern-Simons theory is that it allows us to use
Witten's remarkable result relating Chern-Simons theory and the
Jones polynomial of knot theory \cite{Witten89}. Braiding matrix elements can
be obtained by computing the Jones polynomial of the corresponding
link diagrams \cite{Fradkin98,DasSarma05,Bonderson05}.
For instance, the qubit-flipping property of the process which
takes $\eta_3$ around $\eta_1$ in (\ref{eqn:fourqh}) can be
obtained by evaluating the Jones polynomial (operator) at $q=e^{\pi i/4}$
for the links in Figure \ref{fig:Jones-qubit}.

\begin{figure}[tbh]
 \includegraphics[width=3.25in]{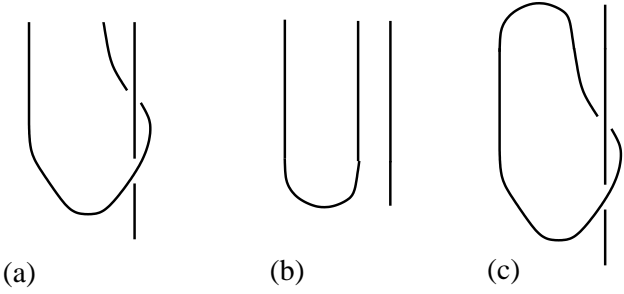}
\caption{(a) A quasiparticle trajectory which winds around a quasiparticle at an antidot
(b) a quasiparticle trajectory which doesn't wind around the antidot; (c)
the matrix element between the states resulting from these two trajectories.}
\label{fig:spacetime-normal}
\end{figure}

What do these pictures mean and how are they evaluated?
They are essentially Feynman diagrams for the topological
interactions of quasiparticles.
The lines in this picture represent isospin $1/2$ quasiparticles.
Since the electric charge is not accounted for in these diagrams,
quasiparticles and quasiholes are identical as far as these diagrams are
concerned. (The electric charge just
contributes Abelian phase factors.) If two of these isospin $1/2$ quasiparticles
fuse to form a spin $0$, then they can annihilate to give the vacuum.
Otherwise, they fuse to form an isospin $1$ particle, which we denote
by a boxed $1$ on the two lines (as in Figure \ref{fig:Jones-qubit}b, for example).
Consider, for instance, the spacetime trajectories of
 the test quasiparticle and the quasiparticle on the antidot in Figure \ref{fig:claudio-dev}.
 They will look like Figure \ref{fig:spacetime-normal}a.
 In this figure, we assume that
 a quasiparticle-quasihole pair is created at $P$ and the quasiparticle goes around the
 antidot. The other trajectory, with which this interferes, is depicted in figure 
 \ref{fig:spacetime-normal}b: in this case, the test quasiparticle does not go around the
 antidot. The interference term between these two processes, i.e. the third term
 in equation \ref{eqn:inter-cond}, is given by the matrix element between these
 two processes, depicted in Figure \ref{fig:spacetime-normal}c.
 It is just Figure \ref{fig:spacetime-normal}b inverted (which corresponds
 to turning a ket into a bra) and then stacked on top of \ref{fig:spacetime-normal}a.
 This matrix element measures the topological charge associated with
 the loop in Figure \ref{fig:spacetime-normal}c.
 
 Hence, to summarize: we specify an initial state $|{\psi_0}\rangle$
 of $n$ qubits by drawing $2n$ incoming lines, which are grouped
 into $n$ pairs. At the bottom of the diagram (the distant past), each pair
 is either created from the vacuum (in which case this qubit is in the state $|0\rangle$)
 or was fused to form an isospin  $1$ particle
 (in which case this qubit is in the state $|1\rangle$).
 The final state $|\psi\rangle$ is obtained by braiding these quasiparticles,
 which is depicted in the figure with the vertical direction on the page being positive time.
 Finally, the matrix element between $|\psi\rangle$
 and another state $|\chi\rangle$ is obtained by inverting the corresponding diagram for
 $|\chi\rangle$ and stacking it on top of the diagram for $|\psi\rangle$, connecting
 each quasiparticle line from $\langle\chi |$ to the corresponding quasiparticle
 line from $|\psi\rangle$.

The matrix element associated with such a diagram is evaluated recursively in the following way.
Each diagram is replaced by the sum of two diagrams both of which have one fewer
crossing according to the rule:
\begin{equation}
  \label{eq:recursion}
  \pspicture[0.4](1.0,1.0)
   \psbezier[linewidth=1.0pt](0.333333333333333,0)(0.333333333333333,0.5)(0.666666666666667,0.5)
  (0.666666666666667,1.0)
  \psbezier[linewidth=1.0pt](0.666666666666667,0)(0.666666666666667,0.333333333333333)
  (0.6,0.4)(0.6,0.45)
  \psbezier[linewidth=1.0pt]
  (0.4,0.55)(0.4,0.6)(0.333333333333333,0.666666666666667)(0.333333333333333,1.0)
  \endpspicture
  \:=\: q^{1/2}{\hskip-0.2cm}
  \pspicture[0.4](1.0,1.0)
   \psbezier[linewidth=1.0pt](0.333333333333333,0)
  (0.333333333333333,0.5)(0.333333333333333,0.5)(0.333333333333333,1.0)
  \psbezier[linewidth=1.0pt](0.666666666666667,0)(0.666666666666667,0.5)
  (0.666666666666667,0.5)(0.666666666666667,1.0)
  \endpspicture
  \; + \; q^{-1/2}{\hskip-0.2cm}
  \pspicture[0.4](1.0,1.0)
  \psbezier[linewidth=1.0pt](0.333333333333333,0)
  (0.333333333333333,0.333333333333333)
  (0.666666666666667,0.333333333333333)(0.666666666666667,0)
  \psbezier[linewidth=1.0pt](0.666666666666667,1.0)
  (0.666666666666667,0.666666666666667)
  (0.333333333333333,0.666666666666667)(0.333333333333333,1.0)
  \endpspicture
\end{equation}
The new diagrams come with corresponding coefficients as indicated
in the above relation. We continue this procedure until all crossings are eliminated.
The projector on the isospin $1$ state is replaced by the explicit expression:
\begin{equation}
  \label{eq:JW1}
  \pspicture[0.4](1.0,1.0)
  \psline[linewidth=1.0pt](0.333333333333333,0)(0.333333333333333,0.3)
  \psline[linewidth=1.0pt](0.333333333333333,0.7)(0.333333333333333,1.0)
  \psline[linewidth=1.0pt](0.666666666666667,0)(0.666666666666667,0.3)
  \psline[linewidth=1.0pt](0.666666666666667,0.7)(0.666666666666667,1.0)
  \pspolygon[linewidth=1.0pt](0.25,0.3)(0.75,0.3)
  (0.75,0.7)(0.25,0.7)
  \uput[u](0.5,0.22){$1$}
  \endpspicture
  \:=\:
  \pspicture[0.4](1.0,1.0)
  \psbezier[linewidth=1.0pt](0.333333333333333,0)
  (0.333333333333333,0.5)(0.333333333333333,0.5)(0.333333333333333,1.0)
  \psbezier[linewidth=1.0pt](0.666666666666667,0)(0.666666666666667,0.5)
  (0.666666666666667,0.5)(0.666666666666667,1.0)
  \endpspicture
  \: - \; \frac{1}{d}{\hskip-0.2cm}
  \pspicture[0.4](1.0,1.0)
  \psbezier[linewidth=1.0pt](0.333333333333333,0)
  (0.333333333333333,0.333333333333333)
  (0.666666666666667,0.333333333333333)(0.666666666666667,0)
  \psbezier[linewidth=1.0pt](0.666666666666667,1.0)
  (0.666666666666667,0.666666666666667)
  (0.333333333333333,0.666666666666667)(0.333333333333333,1.0)
  \endpspicture
\end{equation}
where $d=-q-q^{-1}$.
Finally, every closed loop is given the value
\begin{equation}
  \pspicture[0.4](1.0,1.0)
  \pscircle(0.5,0.5){0.333333333333333}
  \endpspicture
   =\: d
\end{equation}

\section{The effect of adding or deleting bands of material to a topological state}
\label{sec:adding-handles}

In this appendix, we discuss how a $\nu=5/2$ fractional quantum Hall
droplet is modified by topology change, in particular when bands of material
are added or deleted. The basic guideline is that adding a band of material
does not change the state of the system, so it is reversible;
deleting a band of material can cause the state of the system to change
irreversibly.

\begin{figure}[th]
\centerline{\includegraphics[width=3.75in]{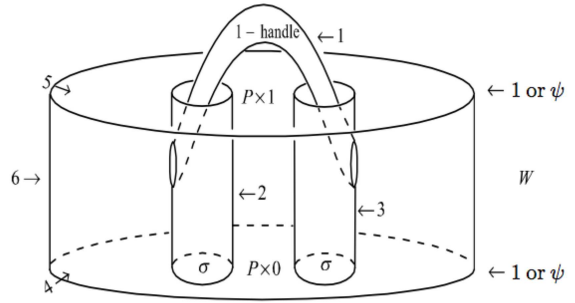}}
\caption{The spacetime history of a process in which
an overpass connecting two antidots is added and then removed. Warning: this history does not embed in $2+1$
dimensions, so the $z-$coordinate in this picture cannot be literally interpreted as time. The handle must move into a fourth direction.}
\label{fig:handle-added}
\end{figure}

In the Bravyi-Kitaev construction discussed in section \ref{sec:Bravyi-Kitaev},
we described a process in which a band of material (the overpass in the middle
picture in Figure \ref{fig:overpass}) is added and then later removed.
The spacetime history of such a process is depicted in Figure
\ref{fig:handle-added}.
In this figure, two $\sigma$ quasiparticles on anti-dots fuse to form $1$ or $\psi$, which is the
topological charge of the boundary. The spacetime history of the quantum Hall
droplet with two anti-dots
is shown as a `pair of pants' legs $P\times I$, where $P$ is the disk with two punctures
and $I$ is the interval of time $[0,1]$.
The spacetime history depicting the addition
and subsequent removal of an overpass connecting the anti-dots is a handle
connecting the two pants legs. The union of this $1$-handle and the spacetime
history of the droplet with two anti-dots is $W=P\times I \cup (1-\text{handle})$.

We would like to check that this procedure leaves invariant the qubit supported by $P$
(rather than, say, applies a phase gate).
In order to do this, we imagine evaluating the Chern-Simons functional integral
on the $3$-manifold $W$. It will give a state in the Hilbert space of the boundary
$\partial W$. However, this boundary is divided into subsurfaces by loops
bounding specified topological charges, so the Hilbert space of the boundary is,
in fact, the tensor product of the Hilbert spaces of the subsurfaces. Hence,
$W$ specifies a vector $\psi_1$ in: $V_{0,0} \otimes V_{0,\s,\s} \otimes V_{0,\s,\s}\otimes V_{\s,\s, x} \otimes  V_{\s,\s,x}^{\ast} \otimes V_{x,x}$, where the factors come from subsurfaces $1, \ldots, 6$ in Figure 22.  The zero label in the first three factors is dictated by the presence of
disks in $W$ capping the boundary of the first component (a cylinder). 
The gluing axiom \cite{Walker91,Turaev00} tells us that removing the $1-$handle determines a canonical isomorphism to $Z(P\times I)$ carrying $\psi_1$ to $\psi_0$ in  $V_{0}^{\ast} \otimes V_{0} \otimes V_{0,\s,\s}\otimes V_{0, \s,\s} \otimes  V_{\s,\s,x} \otimes V_{\s,\s, x}^{\ast} \otimes V_{x,x}$.  After supplying the canonical base vectors $\beta_0^\ast \in V_0^\ast , \beta_{0,\s,\s} \in V_{0,\s, \s}$ and $\beta_{x, x}\in V_{x,x}, \psi_1$ is canonically identified with id $\in$ Hom $(V_{\s,\s,x}) \cong V_{\s,\s, x}^{\ast} \otimes V_{\s,\s,x}$.  Note that no $x-$dependent phase has entered the calculation.  Thus we have proved, in the abstract language of TQFTs, that adding and then breaking a band induces the identity operator on the qubit supported in $P$.

The situation is rather different if, instead, we cut out a band to join the internal punctures and then restore it (i.e. fuse the internal punctures and then separate them).
This is true regardless of whether
we assume that we can use the electric charge of the $\sigma$s
to ensure that each resulting puncture again carries a charge-$\frac{e}{4}$
$\sigma$ after they are fused and split.
Such an operation is depicted in Figure \ref{fig:spacetime-tilted}.
We show below that this operation acts on $P$ by the identity
if $\gamma$ has topological charge $1$. If it has topological charge
$\psi$, however, it is $\sigma_z$. If it is a linear superposition of
these two possibilities, then the applied gate is the same linear combination
of $1$ and $\sigma_z$. The moral, in general, is that operations which add
quantum media (in this case, $\nu=5/2-$ FQHE fluid) are reversible --
simply delete what was previously added, whereas operations which delete
are often irreversible.

This operation can be depicted in the Feynman diagram notation introduced in
appendix \ref{sec:EFT}. It is either Figure \ref{fig:merge}a or \ref{fig:merge}b,
according to the charge of $\gamma$.

\begin{figure}[th]
\centerline{\includegraphics[width=3in]{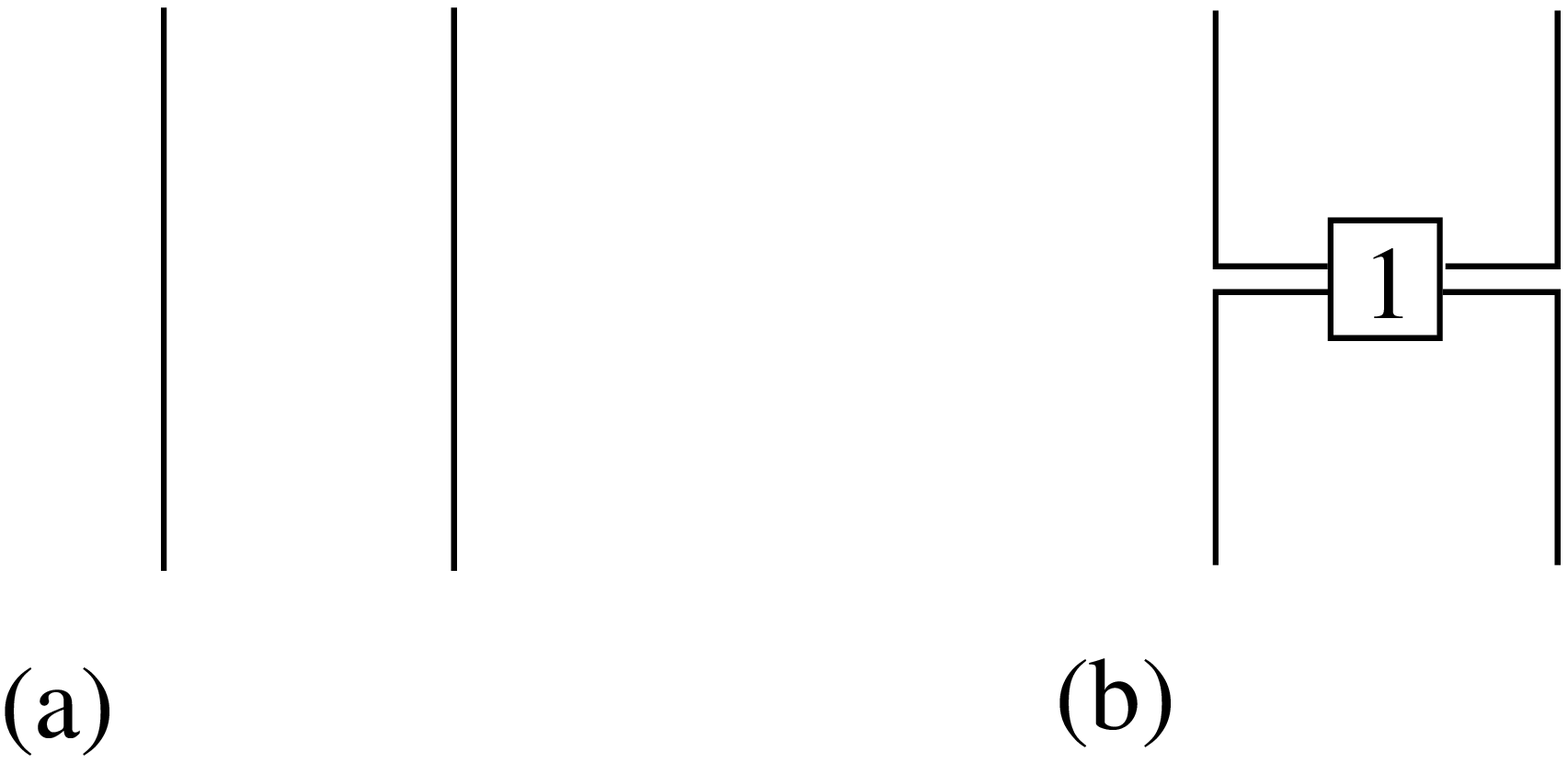}}
\caption{The spacetime history of Figure \ref{fig:handle-added}
is equivalent to the Feynman diagrams above in the notation introduced in
appendix \ref{sec:EFT}.}
\label{fig:merge}
\end{figure}

Clearly, both operators are diagonal in the $(1, \psi)$ basis. Furthermore,
Figure \ref{fig:merge}a is simply the identity acting on the $P-$qubit. This may be verified
by direct computation of the diagonal entries. Using the rules introduced in appendix \ref{sec:EFT},
we $\sqrt{2} \cdot 1$. The strange $\sqrt{2}$ factor is actually $S_{11} = S_{\psi\psi}$ which has entered because we have not rescaled the dual physical Hilbert space by $1/S_{11}$ prior to gluing.
Taking this factor into account, we obtain the identity.

Computing the diagonal entries for the second process,
we obtain $\sigma_z$, as claimed, after taking into account
the correction described above and rescaling by $1/S_{\psi\psi}$.

\end{document}